\documentclass[11pt,a4paper]{article}
\pdfoutput=1
\usepackage{jcappub}

\usepackage{bm}
\usepackage{amsfonts}
\usepackage{subfigure}

\newcommand{\alt}{\mbox{\;\raisebox{.3ex}
  {$<$}$\!\!\!\!\!$\raisebox{-.9ex}{$\sim$}\;}}

\newcommand{\be}{\begin{equation}}
\newcommand{\ee}{\end{equation}}
\newcommand{\bea}{\begin{eqnarray}}
\newcommand{\eea}{\end{eqnarray}}

\begin{document}

%%%%%%%%%%%%%%%%%%%%%%%%%%%%%%%%%%%%%%%%%%%%%%%%%%%%%%%%%%%%%%%%%%%%%%
% Frontpage %%%%%%%%%%%%%%%%%%%%%%%%%%%%%%%%%%%%%%%%%%%%%%%%%%%%%%%%%%
%%%%%%%%%%%%%%%%%%%%%%%%%%%%%%%%%%%%%%%%%%%%%%%%%%%%%%%%%%%%%%%%%%%%%%

%\subheader{\hfill Preprint ....}

\title{Fourier imaging of non-linear structure formation}

\author[a,b]{Jacob Brandbyge,}
\author[a]{Steen Hannestad}

\affiliation[a]{Department of Physics and Astronomy, University of Aarhus, Ny Munkegade 120, DK--8000 Aarhus C, Denmark}
\affiliation[b]{Centre for Star and Planet Formation, Niels Bohr Institute \& Natural History Museum of Denmark, University of Copenhagen, {\O}ster Voldgade 5-7, DK--1350 Copenhagen, Denmark}

\emailAdd{jacobb@phys.au.dk, sth@phys.au.dk}

\abstract{We perform a Fourier space decomposition of the dynamics of non-linear cosmological structure formation in $\Lambda$CDM models.  From $N$-body simulations involving only cold dark matter we calculate 3-dimensional non-linear density, velocity divergence and vorticity Fourier realizations, and use these to calculate the fully non-linear mode coupling integrals in the corresponding fluid equations. Our approach allows for a reconstruction of the amount of mode coupling between any two wavenumbers as a function of redshift. With our Fourier decomposition method we identify the transfer of power from larger to smaller scales, the stable clustering regime, the scale where vorticity becomes important, and the suppression of the non-linear divergence power spectrum as compared to linear theory. Our results can be used to improve and calibrate semi-analytical structure formation models.
}

\maketitle

%%%%%%%%%%%%%%%%%
%%%%%%%%%%%%%%%%%

\section{Introduction}

Cosmological structure formation can in principle be tracked exactly by solving the coupled Einstein-Boltzmann equations for all species. 
For species which are non-relativistic (such as cold dark matter and baryons) it is advantageous to solve the hierarchy of momentum integrated Boltzmann equations, i.e.\ the continuity equation, the Euler equation, and corresponding equations for the higher velocity moments, and using the fact that for non-relativistic species the higher velocity moments are strongly suppressed. For cold dark matter it is a very good approximation to truncate the hierarchy equations and simply solve the continuity and Euler equations.

In linear perturbation theory these equations are easily solved, but on smaller scales ($d \alt 100-200$ Mpc) structures have evolved into the non-linear regime and perturbation theory breaks down. This means that terms which are quadratic in the fluid variables must be taken into account.
Methods based on diagram resummation \cite{Peloso:2016qdr,Floerchinger:2016hja,Senatore:2014via,Carroll:2013oxa,Carrasco:2012cv,Bernardeau:2008fa,Matsubara:2007wj,Pietroni:2008jx,Lesgourgues:2009am,Matarrese:2007wc,Crocce:2005xy} or higher order perturbation theory \cite{Blas:2013aba,Carlson:2009it,Carrasco:2013sva,Rampf:2016wom,Villa:2015ppa,Rampf:2012xb,Scoccimarro:1996se} have been shown to be fairly accurate beyond the scales where linear theory breaks down. However, they are only applicable in the semi-linear regime and fail once the truly non-linear regime is entered. Alternatively, semi-analytic methods such as HALOFIT \cite{Smith:2002dz} use knowledge of the qualitative behaviour of quantities such as the power spectrum in the linear and non-linear regimes as well as a number of fitting parameters to interpolate between the two regimes. These parameters in turn must be calibrated using $N$-body simulations, and using the best current methods an accuracy of a few percent can be achieved for $\Lambda$CDM-like models (see e.g.\ \cite{Takahashi:2012em}).

Even in the simplest possible case of a pure Cold Dark Matter (CDM) density field, the problem of non-linear structure formation is highly challenging. Formally, the problem is to solve the Boltzmann equation for a collisionless fluid in a metric provided by the Einstein equations. On scales much smaller than the horizon this extremely hard problem can be well approximated by Newtonian theory in an expanding FRW background. Since CDM is assumed to be cold, it can be approximated by a non-viscous fluid with $c_s = 0$ and the Boltzmann equation can then be reduced to the continuity and Euler equations, i.e.\ the first two moment-equations of the Boltzmann hierarchy.

A fully non-linear calculation directly in Fourier space requires one to perform a 3-dimensional integral for each wavenumber. The time required to perform structure formation simulations with this method is several orders of magnitude larger than what can currently be considered feasible. The most successful approach to solving these equations is instead to use $N$-body simulations (i.e.\ a coarse grained particle representation of phase-space) and simply solve the Newtonian equation of motion for all particles. This approach can be mainly thought of as a real space solution of the fluid equations, although fast Fourier transforms are used for calculating long-range gravitational forces.

The next generation of large-scale structure surveys such as EUCLID will achieve a measurement of observables like the shear power spectrum to almost cosmic variance precision over a wide range of scales. In order to extract information on the underlying cosmological parameters this in turn requires quantities such as the matter power spectrum to be calculated to 1\% level precision (see e.g.\ \cite{Schneider:2015yka} for a recent discussion). While this is achievable for a few different models it is currently impossible to perform scans over multi-dimensional parameter spaces using $N$-body simulations. The theoretical understanding and numerical implementation of non-linear cosmological structure formation models is therefore significantly challenged by the progress in observational cosmology. To confront this challenge, one needs a deep understanding of the dynamics of cosmological mode coupling and thereby how non-linear structures form in the Universe. 

The purpose of the present study is to gain insights into the structure of the continuity and Euler equations in Fourier space. Using $N$-body simulations we calculate exactly the non-linear source terms in these equations and identify the various regimes, from purely linear structure formation on large scales to the stable clustering regime governing the smallest scales. We propose that these insights can be used to construct new semi-analytic methods for calculating quantities such as the matter power spectrum.

Since we neglect higher moments in the Euler equation, our calculations do not contain the full picture of structure formation at scales close to virialization, $k\sim 1 h/{\rm Mpc}$ at $z\sim 0$. At these scales, anisotropic stress must be taken into account.

The high dimensionality of the problem requires $N$-body simulations to follow the time-evolution, but it is possible at specified redshifts to perform a full non-linear calculation in Fourier space, and thereby probe the dynamics of cosmological mode coupling.

We note that \cite{Pueblas:2008uv} and \cite{Nishimichi:2014rra} also investigate the effect of mode coupling using $N$-body simulations, but in a very different way from the one presented in this paper. 

The paper is structured as follows. In section \ref{sec:equations} we derive the non-linear continuity and Euler equations in Fourier space from relativistic energy-momentum conservation. In section \ref{sec:implementation} we then explain how we have implemented the calculation of the mode coupling integrals in these equations, and in section \ref{sec:results} we present our results. Finally, section \ref{sec:conclusions} contains our conclusions.

%%%%%%%%%%%%%%%%%%%%%%%%%%%%%%%%%%%%%%%%%%%%%%%%%%%%%%%%%%%%%%%%%%%%%%%%%%
\section{The non-linear fluid equations}\label{sec:equations}
%%%%%%%%%%%%%%%%%%%%%%%%%%%%%%%%%%%%%%%%%%%%%%%%%%%%%%%%%%%%%%%%%%%%%%%%%%
The following theoretical outline is performed in the conformal Newtonian gauge \cite{Ma:1995ey}, where deviations from flat space-time are parametrized in terms of the metric potentials $\phi$ and $\psi$, $a$ is the scale factor, and $\tau$ is the proper time. The line element is given by
\begin{equation}
	ds^2 = g_{\mu\nu} dx^\mu dx^\nu = a^2(\tau) [- (1+2\psi)d\tau^2 + (1-2\phi) \delta_{ij} dx^i dx^j].
\end{equation}

%%%%%%%%%%%%%%%%%%%%%%%%%%%%%%%%%%%%%%%%%%%%%%%%%%%%%%%%%%%%%%%%%%%%%%%%%%%%%%%%%
%%%%%%%%%%%%%%%%%%%%%%%%%%%%%%%%%%%%%%%%%%%%%%%%%%%%%%%%%%%%%%%%%%%%%%%%%%%%%%%%%
The relativistic evolution of the energy-momentum tensor, $T^\mu_\nu$, is determined by requiring its covariant derivative to vanish
\begin{equation}
	T^\mu_{\nu;\mu} = \partial_\mu T^\mu_\nu + \Gamma^\lambda_{\lambda\mu} T^\mu_\nu - \Gamma^\lambda_{\mu\nu} T^\mu_\lambda = 0.
	\label{eq:conservation}
\end{equation}

The 10 independent components in the symmetric energy-momentum tensor can be used to define 10 fluid variables, $\delta$, $u^i$, $\delta P/\delta\rho$ and $\Sigma^i_j$, as follows
\begin{align}
	T^0_0 &= - \bar\rho(1+\delta), \nonumber\\
	T^i_0 &= -\bar\rho\left(1+\delta+w+ \frac{\delta P}{\delta\rho}\delta\right)u^i, \nonumber\\
	 T^i_j &= \bar\rho\left(w+\frac{\delta P}{\delta\rho}\delta \right)\delta^i_j +\Sigma^i_j,
\end{align}
where $w=\bar P/\bar \rho$, $\bar\rho$ and $\bar P$ is the average density and pressure, respectively, $\delta^i_j$ is the Kronecker delta and $\Sigma_i^i = 0$.

\subsection{Real space equations}
Below we linearize the equations with respect to the metric perturbations, $\psi$ and $\phi$, since they are very small for all scales relevant to cosmology. The 10 fluid variables are not assumed to be small, and the equations are therefore non-linear in these quantities. After deriving the full general relativistic equations we identify the terms relevant for studying mode coupling for a cold fluid in a Newtonian setting.

\paragraph{The continuity equation in real space and its Newtonian limit}

Setting $\nu = 0$ in Eq.~(\ref{eq:conservation}) and defining the velocity divergence $\theta = \partial_i u^i$ we get the continuity equation
\begin{align}
	\dot\delta =& -(1+w)(\theta - 3\dot\phi) -3\frac{\dot a}{a}\left(\frac{\delta P}{\delta\rho} - w \right) \delta  \nonumber \\
				& -\theta\delta - u^i \partial_i \delta  \nonumber\\
				 &+3\left(1+\frac{\delta P}{\delta\rho}\right)\dot\phi\delta  - \frac{\delta P}{\delta\rho} \theta\delta - u^i \partial_i \left(\frac{\delta P}{\delta \rho}\delta\right)  \nonumber \\
                                & -(\partial_i \psi - 3\partial_i \phi) \left(1+\delta + w + \frac{\delta P}{\delta \rho} \delta \right) u^i.
\end{align}
The terms in the first line are linear terms, the ones in the second line are Newtonian non-linear terms, while the third line includes pressure and non-linear horizon terms. The terms in the last line can be neglected since $(\partial_i \psi - 3\partial_i \phi)u^i = \partial_i( \psi u^i - 3\phi u^i) - (\psi - 3\phi)\theta \ll \theta$. Here we have assumed that the metric potentials are much less than unity, and that, e.g.,  $\partial_i(\psi u^i) \ll \partial_i u^i$, since the general relativistic version of the Poisson equation constrains the relation between spatial derivatives of the metric potentials and the fluid variables.

The non-linear Newtonian limit of the continuity equation is given by
\begin{equation}
	\dot \delta = -(1+\delta)\theta - u^i \partial_i \delta,
	\label{eq:dot_delta}
\end{equation}
where pressure has been neglected, as well as time-derivatives of the metric perturbations.

\paragraph{The Euler equation in real space and its Newtonian limit}
Setting $\nu = i$ in the energy-momentum conservation equations, Eq.~(\ref{eq:conservation}), one arrives at the Euler equation

\begin{align}
	\dot u^i =& -\left[\frac{\dot a}{a}(1-3w) - \dot\psi - 5\dot\phi\right] u^i - \frac{\left[\dot\delta +\dot w + \partial_\tau\left(\frac{\delta P}{\delta\rho}\delta\right)\right]u^i + \delta^{ij} (1+\delta)\partial_j\psi }{1+\delta + w+\frac{\delta P}{\delta\rho}\delta} \nonumber\\
                          &-\frac{1}{\bar\rho\left(1+\delta + w+\frac{\delta P}{\delta\rho}\delta\right)}\left[  \delta^{ik} (\partial_j + \partial_j \psi -3 \partial_j\phi) + \delta^k_j \delta^{il} \partial_l \phi \right]  T^j_k.
      \label{eq:euler1}
\end{align}
As was the case for the continuity equation, we assume that the metric potentials do not vary extremely rapidly, and therefore that the flow velocities are small compared to the speed of light. With this assumption, one can neglect the $\partial_j \phi$ terms in the second line, but one must keep the $\partial_j\psi T^j_k$ term for species with isotropic pressure.

The energy-momentum tensor for a perfect fluid is given by $T^\mu_\nu =P\delta^\mu_\nu +(\rho + P) U^\mu U_\nu$. Since $U^\mu = dx^\mu / \sqrt{-ds^2}$ and $u^i = dx^i/d\tau$, $U^i = u^i / a$ to lowest order in the metric potentials. 

Throughout this work we will assume the Pressureless Perfect Fluid approximation and set $P=0$ (i.e.\ our results will be applicable to any cold dark matter simulation but not to cases where e.g.\ neutrinos are important). The spatial part of $T^\mu_\nu$ can then be written as 
\begin{align}
	\tilde T^i_j =& \bar\rho(1+\delta) u^i u^k \delta_{jk},
\end{align}
and from it we define the intrinsic stress tensor $\sigma^i_j$ as follows
\begin{equation}
	(\rho + P) \sigma^i_j = T^i_j - \tilde T^i_j.
\end{equation}
Note that this definition is different from the stress tensor $\Sigma^i_j$.

The non-linear Newtonian limit of the Euler equation can then be found by neglecting time-derivatives of the metric potentials and terms involving $P$. Using Eq.~(\ref{eq:dot_delta}), Eq.~(\ref{eq:euler1}) then reduces to
\begin{equation}
	\dot u^i = -\frac{\dot a}{a}u^i -u^j \partial_j u^i -\delta^{ij} \partial_j \psi - \frac{\delta^{ik}}{\rho} \partial_j \left( \rho \sigma^j_k \right).
	\label{eq:euler2}
\end{equation}
From the intrinsic stress tensor we can then define divergence and curl quantities as follows
\begin{align}
	q_\theta \equiv &~ \delta^{ik}\partial_i \left[ \frac{1}{\rho} \partial_j \left( \rho \sigma^j_k \right)\right], \\
	q^i_w      \equiv & ~ \varepsilon^{ijk} \partial_j \left[ \frac{1}{\rho} \partial_l \left( \rho \sigma^l_k \right)\right],
\end{align}
where $\varepsilon^{ijk}$ is the Levi-Civita symbol.  Throughout this paper we will use the terms curl and vorticity interchangeably. Applying $\partial_i$ to Eq.~(\ref{eq:euler2}) one finds the evolution equation for the velocity divergence, namely the Euler equation
\begin{equation}
	\dot \theta = - \frac{\dot a}{a}\theta - \partial_i u^j \partial_j u^i - u^i \partial_i \theta - \nabla^2 \psi - q_\theta.
	\label{eq:dot_theta}
\end{equation}

We now define the vorticity of the velocity field, $w^i = (\nabla \times \mathbf{u})^i$, and construct its equation of motion by applying the curl operator, $\varepsilon^{ijk} \partial_j$, to Eq.~(\ref{eq:euler2})
\begin{equation}
	\dot w^i = - \frac{\dot a}{a}w^i - \delta_{km}\varepsilon^{ilm} \partial_l  (u^j \partial_j u^k) - q_w^i.
	\label{eq:dot_w}
\end{equation}
Since the curl of a gradient is zero the equation of motion is not directly sourced by the gravitational field. Vorticity can therefore only be generated by non-linear evolution through derivatives of the term $u^j\partial_j u^i$.

%%%%%%%%%%%%%%%%%%%%%%%%%%%%%%%%%%%%%%%%%%%%%%%%%%%%%%%%%%%%
\subsection{Fourier space equations}
Define the forward, $\mathbf{F}$, and inverse, $\mathbf{F}^{-1}$, Fourier transforms as
\begin{align}
	\delta(\mathbf{k}) &= \mathbf{F}\left[ \delta(\mathbf{x})\right] = \frac{1}{(2\pi)^3}\int d^3 x ~e^{-i\mathbf{k}\mathbf{x}}~\delta(\mathbf{x}), \nonumber\\
	\delta(\mathbf{x}) &= \mathbf{F}^{-1}\left[ \delta(\mathbf{k})\right] = \int d^3 k ~e^{i\mathbf{k}\mathbf{x}}~\delta(\mathbf{k}).
\end{align}

Rewriting a product of variables in real space in terms of their Fourier components can be achieved by using the convolution theorem
\begin{equation}
	\alpha(\mathbf{x})\beta(\mathbf{x}) =\mathbf{F}^{-1}\left\{\mathbf{F}[\alpha(\mathbf{x})]*\mathbf{F}[\beta(\mathbf{x})]\right\},
\end{equation}
with
\begin{equation}
	\alpha(\mathbf{k})*\beta(\mathbf{k}) = \int d^3 k'  \alpha(\mathbf{k}')\beta(\mathbf{k} - \mathbf{k}') = \int d^3 k' \alpha(\mathbf{k} - \mathbf{k}') \beta(\mathbf{k}'),
\end{equation}
i.e., the convolution operator is commutative. Note that the sum of the arguments equals $\mathbf{k}$, i.e.~when multiplying two Fourier modes the resulting wave has a wavelength equal to the sum of the arguments of the two Fourier modes. 

Using a Helmholtz decomposition the velocity field, $\mathbf{u}$, can be split into a curl-free part, $\nabla\times(\nabla\Upsilon)= 0$, and a divergence-free part, $\nabla\cdot(\nabla\times\mathbf{A})= 0$, as follows $\mathbf{u} = -\nabla \Upsilon + \nabla\times \mathbf{A}$. Using the definitions of $\theta$ and $\mathbf{w}$ this can also be written as $\mathbf{u} = \nabla^{-2}\nabla\theta + (\nabla\times)^{-2} \nabla \times \mathbf{w}$, where $\nabla^2 = \delta^{ij}\partial_i \partial_j$ is the comoving Laplace operator.
The inverse Laplacian, $\nabla^{-2}$, applied to the exponential, $e^{i\mathbf{k}\mathbf{x}}$, has the property $\nabla^{-2} e^{i\mathbf{k}\mathbf{x}} = - k^{-2} e^{i\mathbf{k}\mathbf{x}}$, since $\nabla^2 e^{i\mathbf{k}\mathbf{x}} = - k^2 e^{i\mathbf{k}\mathbf{x}}$,
where $k^2 = \delta^{ij} k_i k_j$. Since $\mathbf{w}$ is divergence free, $\nabla\times\nabla\times\mathbf{w} = - \nabla^2 \mathbf{w}$, so that $(\nabla\times)^{-2} (e^{i\mathbf{k}\mathbf{x}} \mathbf{w}) = k^{-2} (e^{i\mathbf{k}\mathbf{x}} \mathbf{w})$.

\paragraph{The continuity equation in Fourier space}
Defining $I_\delta = \theta \delta + \mathbf{u}\cdot \nabla\delta$,  and using that the Fourier coefficient of $\mathbf{u(\mathbf{x})}$ is given by $\frac{i}{k^2}[-\theta(\mathbf{k})\mathbf{k} + \mathbf{k}\times\mathbf{w}(\mathbf{k})]$, we get
\allowdisplaybreaks
\begin{align}
        \mathbf{F}[I_\delta(\mathbf{x})] =&~ \mathbf{F}[\theta(\mathbf{x})]*\mathbf{F}[\delta(\mathbf{x})] +\mathbf{F}[\mathbf{u}(\mathbf{x})\cdot] * \mathbf{F}[\nabla\delta(\mathbf{x})]       \nonumber\\
		      =& ~ \int d^3k' \theta(\mathbf{k}')\delta(\mathbf{k}'') - \int d^3k' \frac{1}{k'^2} [-\theta(\mathbf{k}') \mathbf{k}'+ \mathbf{k}' \times \mathbf{w}(\mathbf{k}')] \cdot \mathbf{k}'' \delta(\mathbf{k}'')\nonumber\\
     		      =& ~ \int d^3k' \left[\mathbf{k}\cdot\mathbf{k}' \theta(\mathbf{k}')  - \mathbf{k}''\times \mathbf{k}' \cdot \mathbf{w}(\mathbf{k}')\right] \frac{\delta(\mathbf{k}'')}{k'^2},
\end{align}
where $\mathbf{k}''\equiv \mathbf{k-k'}$.

The Newtonian Fourier space version of the continuity equation, Eq.~(\ref{eq:dot_delta}), is then given by
\begin{equation}
	\dot\delta(\mathbf{k}) = - \theta(\mathbf{k})  - I_\delta(\mathbf{k}), 
\end{equation}
where we define $I_\delta(\mathbf{k}) \equiv I_{\theta\delta}(\mathbf{k}) + I_{w\delta}(\mathbf{k})$, with the subscripts referring to the fluid variables in the integral. The individual components are given by
\begin{equation}
	I_{\theta\delta}(\mathbf{k}) =  \int d^3k' \frac{\mathbf{k}\cdot\mathbf{k}'}{k'^2} \theta(\mathbf{k}')  \delta(\mathbf{k}''),
	\label{eq:I_td}
\end{equation}
and
\begin{equation}
	I_{w\delta}(\mathbf{k}) = - \int d^3k' \frac{\mathbf{k}''\times \mathbf{k}'}{k'^2} \cdot \mathbf{w}(\mathbf{k}') \delta(\mathbf{k}'').
       \label{eq:I_wd}
\end{equation}

Since the convolution operator is commutative we could just as well interchange $\mathbf{k}'$ and $\mathbf{k}''$ and thereby get a different appearance for the integrand of $I_\delta(\mathbf{k})$. The two different ways of writing the integrand gives the same value of the integral, but specific features in the integrand will appear at different values of $\mathbf{k}'$ in the two different representations. But, importantly, features at a specific value of $\mathbf{k}'$ measured relative to the argument of $\delta$ is an invariant. We will use this fact below to attach physical meaning to various parts of the integrand.

\paragraph{The Euler equation in Fourier space}
The Euler equation in Fourier space is found by Fourier transforming Eq.~(\ref{eq:dot_theta})
\begin{equation}
	\dot\theta(\mathbf{k}) = - \frac{\dot a}{a}\theta(\mathbf{k}) + k^2\psi(\mathbf{k}) -  q_\theta(\mathbf{k}) - I_\theta(\mathbf{k}),
\end{equation}
with $I_\theta(\mathbf{k}) = I_{\theta\theta}(\mathbf{k})+I_{\theta\mathbf{w}}(\mathbf{k})+I_{\mathbf{w}\mathbf{w}}(\mathbf{k})$.

$I_{\theta\theta}$ is found from the following considerations. The contribution from the real space Euler equation to $I_{\theta\theta}$ is symmetric under exchange of the two $\theta(\mathbf{x})$ variables. The convolution theorem breaks this symmetry, but it can be recovered by applying the convolution theorem twice, with the roles of the two $\theta$s exchanged, and then taking the average of the result. Finally, one arrives at a Fourier space version that is symmetric under exchange of the arguments of the two $\theta$s, namely $\mathbf{k'}$ and $\mathbf{k''}$
\begin{equation}
	I_{\theta\theta}(\mathbf{k}) = \frac{1}{2}k^2\int d^3k' \frac{\mathbf{k}'\cdot\mathbf{k}''}{k'^2 k''^2}\theta(\mathbf{k}')\theta(\mathbf{k}'').
	\label{eq:I_tt}
\end{equation}
In the two cases mentioned above, $\frac{1}{2}k^2$ should be replaced with $k^2 - \mathbf{k} \cdot \mathbf{k}' $ and $\mathbf{k} \cdot \mathbf{k}'$. The version with $\frac{1}{2}k^2$ is the one quoted in \cite{Jain:1993yk, Jain:1993jh, Audren:2011ne, Pueblas:2008uv,Bernardeau:2001qr}. Note that the different ways of writing the integrand introduces an ambiguity in attaching physical meaning to the non-linear contribution from a given value of $\mathbf{k}'$ or $\mathbf{k''}$, but that the version in Eq.~(\ref{eq:I_tt}) is the only one that preserves the symmetry of the real space Euler equation.

The mixed term, with a kernel symmetry broken by the explicit dependence on $\mathbf{k}$, is given by
\begin{equation}
	I_{\theta\mathbf{w}}(\mathbf{k}) = -\int d^3k' \frac{(\mathbf{k+k''})\cdot \mathbf{k'}}{k'^2 k''^2}\theta(\mathbf{k}') \mathbf{k}'\times \mathbf{k}'' \cdot \mathbf{w}(\mathbf{k}'') ,
      \label{eq:I_tw}
\end{equation}
and finally the symmetric curl-curl term is 
\begin{equation}
	I_{\mathbf{w}\mathbf{w}}(\mathbf{k}) = \int d^3k' \frac{ \mathbf{k}''\times \mathbf{k}' \cdot \mathbf{w} (\mathbf{k}')}{k'^2}  
	 	\frac{\mathbf{k}' \times \mathbf{k}'' \cdot \mathbf{w} (\mathbf{k}'')}{k''^2}.
       	\label{eq:I_ww}
\end{equation}

%%%%%%%%%%%%%%%%%%%%%%%%%%%%%%%%%%%%%

\section{Numerical implementation details}\label{sec:implementation}
\subsection{Cosmology and simulations}

In all cases we perform our calculations for a flat cosmology with $\Omega_m=0.3$, $h=0.7$, $n_s=1$, $A_s=2.3\cdot 10^{-9}$, and $\Omega_b=0.05$ (our results do not depend on this specific choice and could equally well have been performed with e.g.\ Planck 2015 best-fit parameters \cite{Ade:2015xua}).
The evolution of density and velocity fluctuations in linear theory is followed using CAMB \cite{Lewis:2002ah}, and the $N$-body particle initial conditions are found from a weighted sum of the CDM and baryon transfer functions. We have used the initial conditions generator first described in \cite{Brandbyge:2008js}, and our pure CDM $N$-body simulations evolved with \textsc{gadget}-2 \cite{Springel:2005mi} are listed in Table~\ref{table:nbody_sims}. The initial conditions are created with either the Zel'dovich Approximation \cite{Zeldovich:1969sb} or with a correction added from second order Lagrangian perturbation theory (2LPT) \cite{Crocce:2006ve, Bouchet:1994xp, Scoccimarro:1997gr}.

\begin{table}[t]
    \begin{center} 
        \begin{tabular}{c c c c c} 
           \hline%\hline 
           Sim & $R_{\rm box}$ & $N_{\rm part}$ & $z_i$ & ZA/2LPT\\
          \hline
          A & 64     & $512^3$   & 199 & ZA\\  
          B & 64     & $512^3$   & 49   & ZA\\  
          C & 64     & $512^3$   & 49   & 2LPT\\  
          D & 256   & $512^3$   & 49   & ZA \\
          E & 1024 & $512^3$   & 49   & ZA\\
          F & 4096 & $512^3$   & 49   & ZA\\
          G & 256   & $1024^3$ & 49   &  ZA\\          
         \hline
      \end{tabular}
      \end{center}
          \caption{The table shows parameters and initialization methods for the $N$-body simulations used in this work. $R_{\rm box}$ denotes the box size in units of ${\rm Mpc} / h$, $N_{\rm part}$ denotes the number of $N$-body particles, $z_i$ is the initial $N$-body starting redshift, where the initial conditions are generated with either the Zel'dovich Approximation (ZA) or with an added term from second order Lagrangian perturbation theory (2LPT).} 
      \label{table:nbody_sims} 
\end{table}

\subsection{Assignment schemes}
When calculating the power spectra of real space fields, it is necessary to assign $N$-body particle masses and velocities to a grid. There are various ways of performing this assignment. A deconvolved Clouds-in-cell (CIC) algorithm works well for the density field, but this method has severe drawbacks for the velocity field. Assume that there are no $N$-body particles in the 8 cubes surrounding a given grid point. For the density field, this will result in $\delta(\mathbf{x}) = -1$, which can be considered a fair estimate of the ``real'' density. But the velocity field will be assigned the value 0, which is a very poor estimate for, e.g., the large-scale bulk flows in voids. These missing values will significantly affect the velocity power spectrum statistics, leading to less power on large scales, and more power on small scales. With the CIC scheme, this error can only be reduced by using a very coarse assignment grid. 

Other assignment schemes can also be used, e.g., the Delaunay tesselation was used in \cite{Pueblas:2008uv,Jennings:2012ej}. In this paper we use the adaptive smoothing length interpolation scheme of \cite{Monaghan:1985} to calculate velocity field statistics. In this method each $N$-body particle is assigned a smoothing length calculated as the distance to its 33rd nearest neighbour ($\pm 2$). This smoothing length is then used to interpolate the particle velocities to all grid points within its smoothing length. As a result, the velocity field is well sampled, also in very low density void regions.

We have investigated the effect of varying the number of neighbours used to find the smoothing length, from 33 to 22 and 66, and found no noticeable effect. Since the smoothing length interpolation kernel varies in space, the interpolation kernel is not deconvolved. The velocity field is found on a $1024^3$ grid, but in our subsequent mode coupling calculations we only use the largest scale modes from a $128^3$ sub-grid. The effect of any deconvolution should be very small on this sub-grid.

For definiteness, we use the deconvolved CIC method to calculate the density field Fourier modes, and the smoothing length method to calculate velocity divergence and curl Fourier modes. In all cases, the modes are calculated on a $1024^3$ regular grid. The divergence and curl quantities are calculated in $k$-space from the velocity Fourier modes.

%%%%%%%%%%%%%%%%%%%%%%%%%%%%%%%%%%%%%%%%%%%%%%%%%%%%%%%%%%%%%%%%%%%%%%%%%%%%%%%%%%%%%%%%%%%%%%%
\subsection{A correlation measure}
The effect of the mode coupling integrals must be measured relative to the linear sourcing terms in the fluid equations. $I_\delta(\mathbf{k})$ should therefore be compared to the sign of $\theta(\mathbf{k})$, and $I_\theta(\mathbf{k})$ should be contrasted with the sign of $-\psi(\mathbf{k}) \propto \delta(\mathbf{k})$. Defining $i_\alpha(\mathbf{k,k'})$ from $I_\alpha(\mathbf{k}) =\int d^3k' i_\alpha(\mathbf{k,k'})$, we construct the following statistics as a measure of the amount of mode coupling between different scales
\begin{equation}
	i_\alpha^\beta(k, k') =\frac{\int d\Omega  s^\beta(\mathbf{k})  \cdot   \int d\Omega' \tilde{i}_\alpha(\mathbf{k}, \mathbf{k}')}{2\int d\Omega},
	\label{eq:avg1}
\end{equation}
where 
\begin{equation}
	s^\beta(\mathbf{k}) = \left (  \begin{array}{c}  {\rm sgn}(\beta_{re}(\mathbf{k})) \\ {\rm sgn}(\beta_{im}(\mathbf{k})) \end{array}\right),
\end{equation}
and
\begin{equation}
	\tilde{i}_\alpha(\mathbf{k,k'}) = \left (  \begin{array}{c} i_\alpha^{re}(\mathbf{k,k'}) \\ i_\alpha^{im}(\mathbf{k,k'}) \end{array}\right),
\end{equation}
where $\beta$ can be either of $\delta$, $\theta$ and $w$.

It is the quantity in Eq.~\ref{eq:avg1} which is presented in the figures in this paper. The correlation measure quantifies the average correlation between two different wavenumbers in the mode coupling integrals found in the continuity and Euler equations.

\subsection{Practical calculation of the mode coupling integrals}
In Eq.~(\ref{eq:avg1}) one must perform a sum over $\Omega'$ and an average over $\Omega$. Since the $\delta(\mathbf{x})$ and $\theta(\mathbf{x})$ fields are real, the corresponding complex Fourier coefficients are packed in the volume with $k_z \ge 0$. For $k_z = 0$ and $k_z=k_N$, where $k_N$ is the Nyquist frequency of the grid, redundant coefficients are stored. 

Now, the deconvolution theorem requires one to perform an integration over all of $k$-space, and not just the part with $k_z \ge 0$. This entails that $\int d\Omega'$ must be performed over all angles. The $\int d\Omega$ integral then estimates an average quantity for a given value of $k$. In this case it is required to perform an isotropic sampling, but this does not require that $\int d\Omega$ should be performed over all angles. Only one half of the angles must be sampled, and we choose the ones with $k_z \ge 0$ for simplicity. The sampling with $k_z < 0$ gives information that is related to the information for $k_z \ge 0$ through complex conjugation.

In practice the integration over $\Omega$ is shifted into one over $\Omega''$ since $\mathbf{k}''$ is an argument of the perturbation variables in the mode coupling integrals in the continuity and Euler equations. 

To create a binned version in $k$ and $\tilde k$, $\tilde k\in(k',~k'')$, performed by the operator $\mathcal{B}$, see below, we need to perform a 6-dimensional integration. We use the definition that $'()'$ around integration limits means that redundant symmetry elements must be removed when performing the integral. To avoid cumbersome notation, the integration limits are denoted with $N$, the integer corresponding to the Nyquist frequency, $k_N$.

The practical implementation of the integrator uses the following rewrite of the 6-dimensional integral

\allowdisplaybreaks
\begin{align}
		i_\alpha^\beta(k, \tilde{k}) = &\int_{-\infty}^\infty dk_x' \int_{-\infty}^\infty dk_y' \int_{-\infty}^\infty dk_z' \int_{-\infty}^\infty dk_x'' \int_{-\infty}^\infty dk_y'' \int_{0}^\infty dk_z'' \nonumber \\
		&~\mathcal{B} \left\{ \tilde{i}_\alpha (\mathbf{k'},\mathbf{k''}) \cdot s^\beta(\mathbf{k'' + k'})\right\}\nonumber \\
		 \rightarrow &\int_{-{\rm N}+1}^{\rm N} dk_x' \int_{-{\rm N}+1}^{\rm N} dk_y' \int_{({0})}^{({\rm N})} dk_z' \int_{-{\rm N}+1}^{\rm N} dk_x'' \int_{-{\rm N}+1}^{\rm N} dk_y'' \int_{({0})}^{({\rm N})} dk_z'' 	\nonumber\\
		 &\mathcal{B} \left\{ \tilde{i}_\alpha(\mathbf{k'},\mathbf{k''}) \cdot s^\beta(\mathbf{k''+k'}) + \tilde{i}_\alpha(\mathbf{-k'},\mathbf{k''}) \cdot s^\beta(\mathbf{k''-k'}) \right\}.
		 \label{eq:practical}
\end{align}
In the above expression $\mathcal{B}$ assigns the first term to $\mathbf{k = k''+k'}$ and the second term to $\mathbf{k = k''-k'}$, and can bin in either $\tilde{k}=k'$ or $\tilde{k}=k''$. For the continuity equation $\alpha \in \{\theta\delta,w\delta\}$ and $\beta=\theta$, whereas $\alpha \in \{\theta\theta,\theta w, ww\}$ and $\beta=\delta$ when calculating the amount of correlation in the Euler equation. Note however that in this paper we do not calculate the $\alpha \in \{\theta w, ww\}$ terms, since they are subdominant on the scales shown.

\subsection{Convergence of results}
We have compared the two simulations with $512^3$ and $1024^3$ $N$-body particles in a $256 {\rm Mpc}/h$ box. The difference in $i_{\theta\delta}^\theta(k'';k)$ is smaller than the thickness of the lines in the figures presented in the following section.

For a given value of the one dimensional grid size, $N$, the Fourier modes are sampled isotropically out to $\tilde{k}_I = \frac{2\pi}{R_{\rm box}}\frac{N}{2}$ for $\tilde{k} \in \{k',k''\}$, i.e. the arguments of the Fourier variables. For $N=128$ and $R_{\rm box} = 64 {\rm Mpc} / h$ this gives $\tilde{k}_I = 2\pi$. This means that, e.g., $i_\alpha^\beta(k'';k)$ is only exactly sampled for $k''<2\pi - k$. This limit is very conservative and the accuracy only decreases slowly with $k''$. 

To assess this issue, we have compared the results for the $64 {\rm Mpc}/h$ box for grid sizes of $64^6$ and $128^6$ at $z=0$. We have found that the $64^6$ and $128^6$ cases give results that deviate by $\sim$ a line thickness in the figures in the next section.

%%%%%%%%%%%%%%%%%%%%%%%%%%%%%%%%%%%%%%%%%%%%%%%%%%%%%%%%%%%%%%%%%%%%%%%%%%%%%%%%%%%%%%%%%%%%%%%
\section{Results}\label{sec:results}
%%%%%%%%%%%%%%%%%%%%%%%%%%%%%%%%%%%%%%%%%%%%%%%%%%%%%%%%%%%%%%%%%%%%%%%%%%%%%%%%%%%%%%%%%%%%%%%
 \subsection{The physical picture and its power spectra statistics}

 \begin{figure}[t]
 \vspace*{-7.7cm}
\begin{center}
 \hspace*{-1.8cm}
\includegraphics[width=18.2cm]{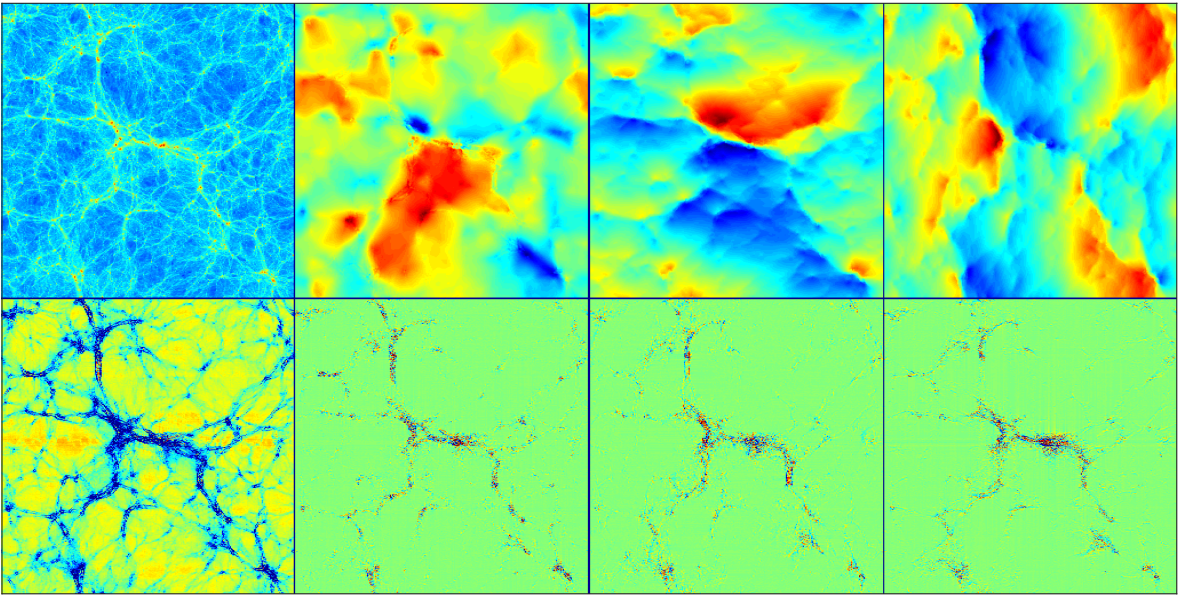}
\end{center}
\vspace*{-8.0cm}
\caption{The figure displays (top row) $\delta$, $v_x$, $v_y$, $v_z$, (bottom row) $\theta$, $w_x$, $w_y$ and $w_z$ (taken from simulation G). The side lengths of each image are $256 {\rm Mpc}/h$, and the thickness of each image is $10 {\rm Mpc}/h$. The projection of the images is along the x-dimension, and this leads to a typologically different image for $v_x$ as compared to $v_y$ and $v_z$.}
   \label{fig:fields}
\end{figure}
Fig.~\ref{fig:fields} shows a slice of the $N$-body simulation volume. The figure displays $\delta$, $\mathbf{v}$, $\theta$ and $\mathbf{w}$. Several things are directly visible. First, the velocity field changes direction perpendicular to the largest filaments, which indicates that $N$-body particles are streaming into the filaments. Second, the velocity field does not display any significant small-scale structure, which mathematically can be understood from the linear theory relation $v \propto \delta / k$. Physically, this means that the gravitational field of the smaller clusters and filaments is too small to significantly alter the velocity of nearby structures. These structures are instead moved by fluctuations on even larger scales.

The velocity divergence field traces the density field exactly in linear theory. But in non-linear theory, the divergence falls behind the density field. Physically this can be understood as follows: The larger scales, $k\sim 0.1 h / {\rm Mpc}$, are dominated by voids, with mass flowing out of these voids today. As the voids become more empty, less mass can flow out of them, a physical fact which is not captured in linear theory, and explains why the linear theory divergence power spectrum is higher than its non-linear counterpart. At smaller scales, $k\gtrsim 0.5 h / {\rm Mpc}$, this effect happened at higher redshift. So even though these scales are dominated by clusters at low redshift, the extra suppression of the divergence power spectrum in non-linear theory is a consequence of mass flowing out of under-density regions at higher redshift.

The vorticity shows virtually no large scale perturbations, but is solely concentrated inside the largest clusters and filaments. Mathematically this can be explained by the fact that the vorticity evolution equation, Eq.~(\ref{eq:dot_w}), is sourced by non-linear perturbations. The vorticity changes sign (from blue to red), over a short distance scale, which shows that the clusters are rotating. Using perturbation theory \cite{Cusin:2016zvu} finds the same shape as we do for the vorticity power spectrum on larger scales, while \cite{Pueblas:2008uv} also predict the turn-over in the vorticity power spectrum.

%%%%%%%%%%%%%%%%%%%%%%%%%%%%%%%%%%%%%%%%%%%%%%%%%%%%%%%%%%%%%%%%%%%%%%%%%%%%%%%%%%%%%%%%%%%%%%%
 \begin{figure}[t]
  \vspace*{-4.0cm}
\begin{center}
\hspace*{-1.0cm}
\includegraphics[width=18cm]{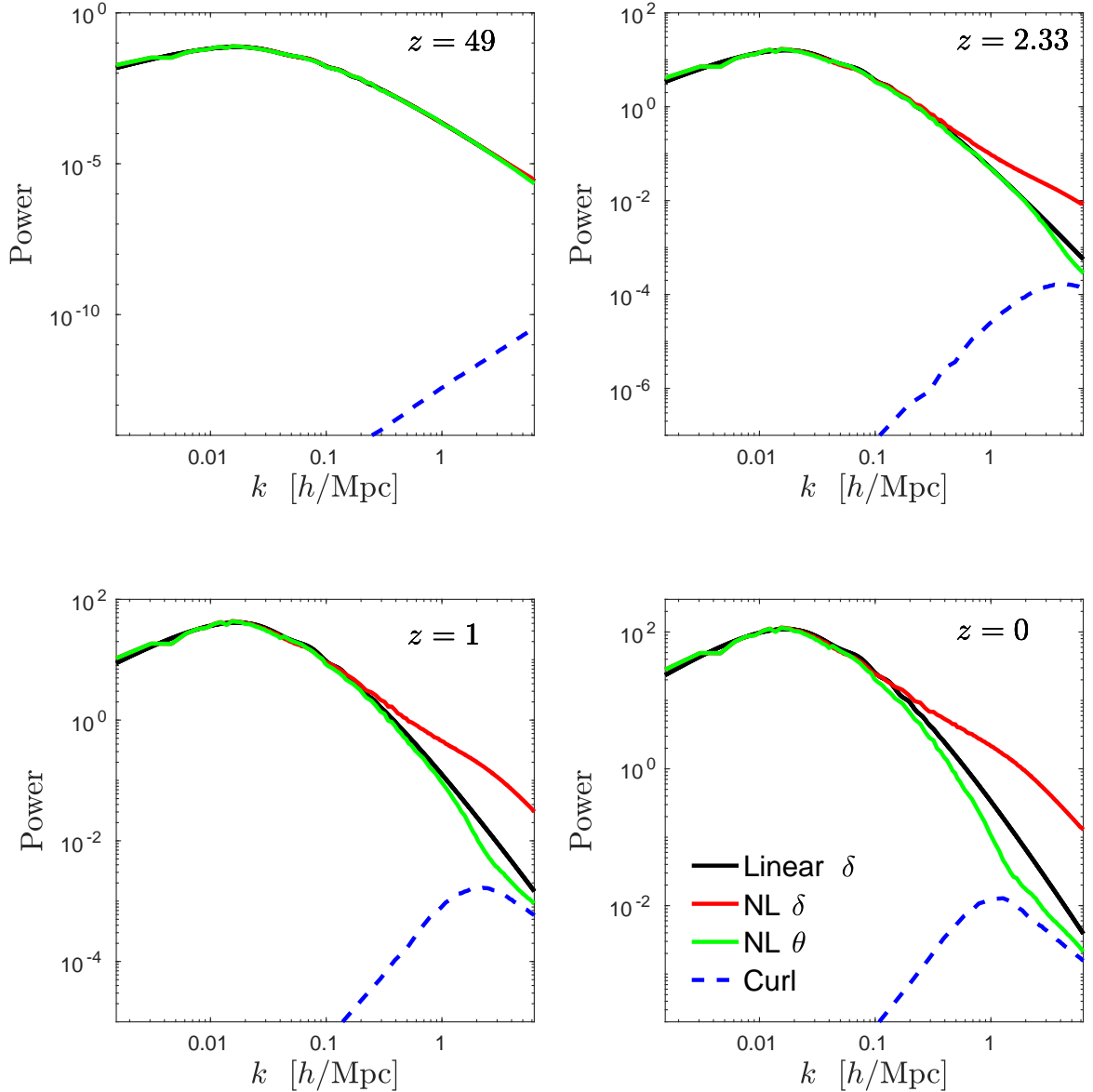}
\end{center}
 \vspace*{-4cm}
\caption{The figure displays the $\delta$ (linear and non-linear), $\theta$ and $\mathbf{w}$ (curl) power spectra at 4 different redshifts. The power spectra are synthesized from the simulations B, D, E and F, see Table~\ref{table:nbody_sims}.}
   \label{fig:power}
\end{figure}

The physical picture from Fig.~\ref{fig:fields} is captured by the power spectra statistics shown in Fig.~\ref{fig:power}. This figure shows the linear theory power spectra for $\delta$ ($\propto \theta$ in linear theory), as well as the non-linear $P_\delta$, $P_\theta$ and $P_w$ power spectra created by combining spectra from 4 different simulations (B, D, E and F). The divergence power spectra have been normalized so that they match the density power spectrum in each image at the largest scale simulated. The divergence and vorticity power spectra are directly comparable. $P_\delta$ is found using the deconvolved CIC mass assignment scheme, whereas $P_\theta$ and $P_w$ are found by using the adaptive smoothing length (SL) method of \cite{Monaghan:1985}. Since the shape of $P_\delta$ and $P_\theta$ are virtually identical at $z=49$, this shows that the SL method, and thereby the velocity power spectra are accurate on the scales shown.

At lower redshift, the non-linear divergence power spectra falls below the linear one. This effect begins at small scales, but as time progresses, it can be seen on much larger scales as well. At $z=0$ the linear and non-linear divergence power spectra only become similar for $k \lesssim 0.02 h/{\rm Mpc}$.

The vorticity power spectrum is completely negligible at large scales and high redshift, and only becomes comparable to $P_\theta$ for $k \gtrsim 1h/{\rm Mpc}$ at $z \lesssim 0.5$. The vorticity power spectra peaks at $k\sim 1.5 h/{\rm Mpc}$ today, at $k\sim 2.5 h/{\rm Mpc}$ for $z=1$ and $k\sim 5 h/{\rm Mpc}$ at $z=2.33$. It is noticeable that the divergence power spectrum has a kink in its slope at these same scales and redshifts. 

Note that the curl power spectrum is non-zero at the $N$-body starting redshift, $z=49$. This initial curl field is generated by the Zel'dovich Approximation.

It is worth noting that the $\delta$ power spectrum changes slope at $k=1.5h/{\rm Mpc}$ at $z=0$. This signifies the stable clustering regime. Finally, at $z=0$ the non-linear density power spectrum lies below linear theory for $k\sim 0.02-0.1 h / {\rm Mpc}$. This suppression of density power is driven by the non-linear suppression in the velocity divergence.

%%%%%%%%%%%%%%%%%%%%%%%%%%%%%%%%%%%%%%%%%%%%%%%%%%%%%%%%%%%%%%%%%%%%%%%%%%%%%%%%%%%%%%%%%%%%%%%
\subsection{Initial mode coupling}
 \begin{figure}[t]
   \vspace*{-5.8cm}
\begin{center}
\hspace*{-0.5cm}
\includegraphics[width=17cm]{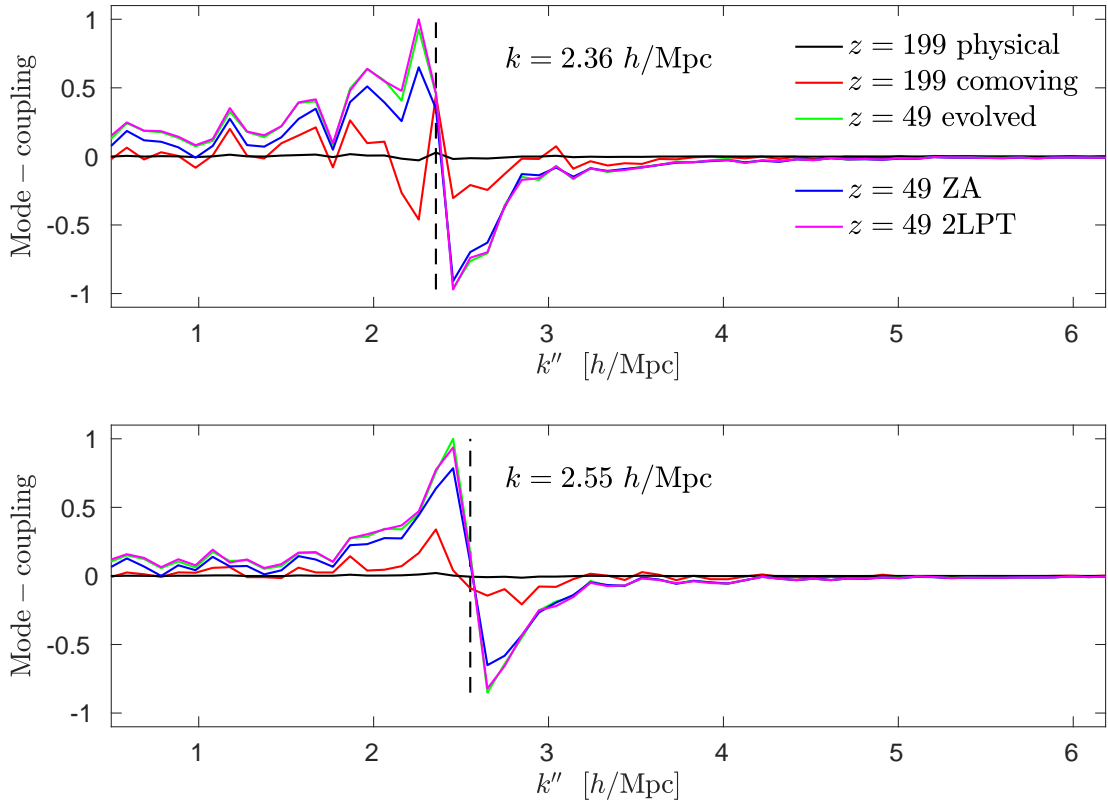}
\end{center}
  \vspace*{-5.7cm}
\caption{The figure shows mode coupling from 3 different simulations. One (simulation A) initilized at $z=199$, black and red lines, where the red line is scaled relative to the black line by the square of the linear growth factor between $z=199$ and $z=49$. The green line shows this simulation evolved until $z=49$. The blue and magenta lines show the amount of mode coupling generated by ZA (simulation B) and 2LPT (simulation C), respectively, from initial conditions created at $z=49$. The vertical dashed line indicates the position of $k$.}
   \label{fig:slices_init}
\end{figure}

Fig.~\ref{fig:slices_init} shows $i_{\theta\delta}^\theta(k'';k)$, Eq.~(\ref{eq:practical}), i.e.~a function of $k''$ for specific values of $k$. The amplitude in each figure is normalised in such a way that the maximum positive amplitude is unity. Results from 3 different simulations are shown. One (simulation A) initialized with ZA at $z=199$ (black and red lines) and then evolved until $z=49$ (green line), another (simulation B) initialized with ZA at $z=49$ (blue line) and finally a simulation (C) initialized with 2LPT at $z=49$ (magenta line). Focusing on the $z=49$ results, it can in all cases be seen that power is moved from larger scales to smaller scales relative to a particular mode ($k$). There is in fact a very sharp transition at $k$ (the value of $k$ is shown with the vertical dashed line). This interpretation is based on the physical understanding, that the non-linear terms in the continuity equation are driven by collapsing density perturbations, and therefore that a physical understanding of the various terms in the mode coupling integral should be based on a binning in the argument of $\delta$, namely $k''$.
 
In the upper panel this effect is not significantly present at $z=199$ due to noise, whereas the power transfer is present in the lower panel. But in both cases the power transfer from larger to smaller scales is a pattern that evolves during the $N$-body simulation. For this to occur, a certain fraction of modes must have an initial pattern that moves power from large to small scales, otherwise the coherent buildup of this pattern over all modes will not evolve over time.
 
The transfer of power from large to small scales can also be generated by the ZA alone, by applying it to a linear theory transfer function and uncorrelated random numbers. This means that the Zel'dovich Approximation by itself generates mode coupling. Adding 2LPT to ZA leads to additional mode coupling, and it can be seen that this additional mode coupling is reproduced by the simulation initialized at $z=199$ and evolved until $z=49$.

%%%%%%%%%%%%%%%%%%%%%%%%%%%%%%%%%%%%%%%%%%%%%%%%%%%%%%%%%%%%%%%%%%%%%%%%%%%%%%%%%%%%%%%%%%%%%%%
\begin{figure}[h!]  \vspace*{-3.5cm} \begin{center} \hspace*{-0.5cm} \includegraphics[width=16cm]{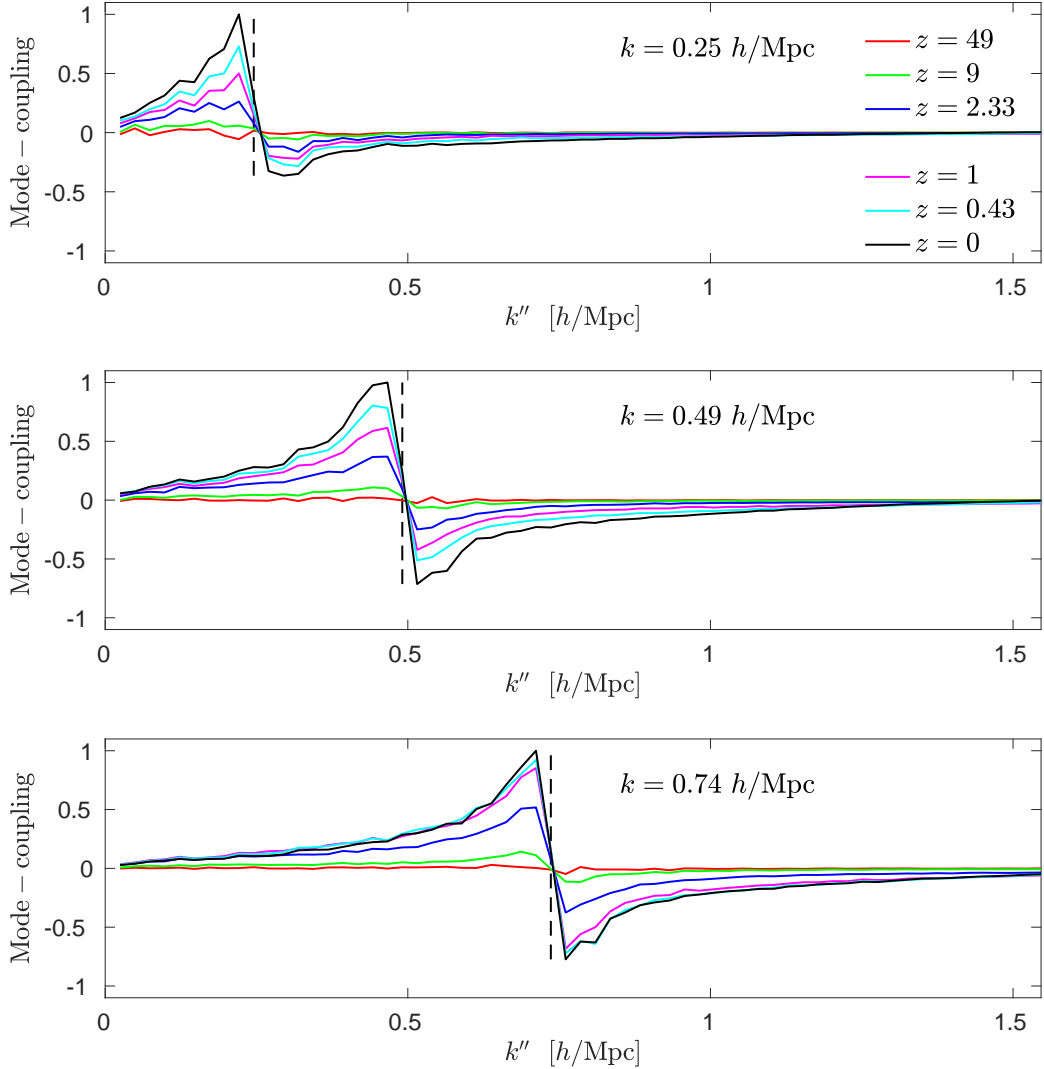} \end{center}  \vspace*{-3.3cm}                              
                              \caption{The figure shows the amount of mode coupling in the continuity equation, $i_{\theta\delta}^\theta(k'';k)$ between $\theta$ and $\delta$, as a function of $k''$ for specific values of $k$. The figure is based on simulation D, see Table~\ref{table:nbody_sims}.}
                              \label{fig:slices46}
\end{figure}

\subsection{The redshift evolution of mode coupling}
\begin{figure}[h!]  \vspace*{-3.5cm}\begin{center}\hspace*{-0.5cm}\includegraphics[width=16cm]{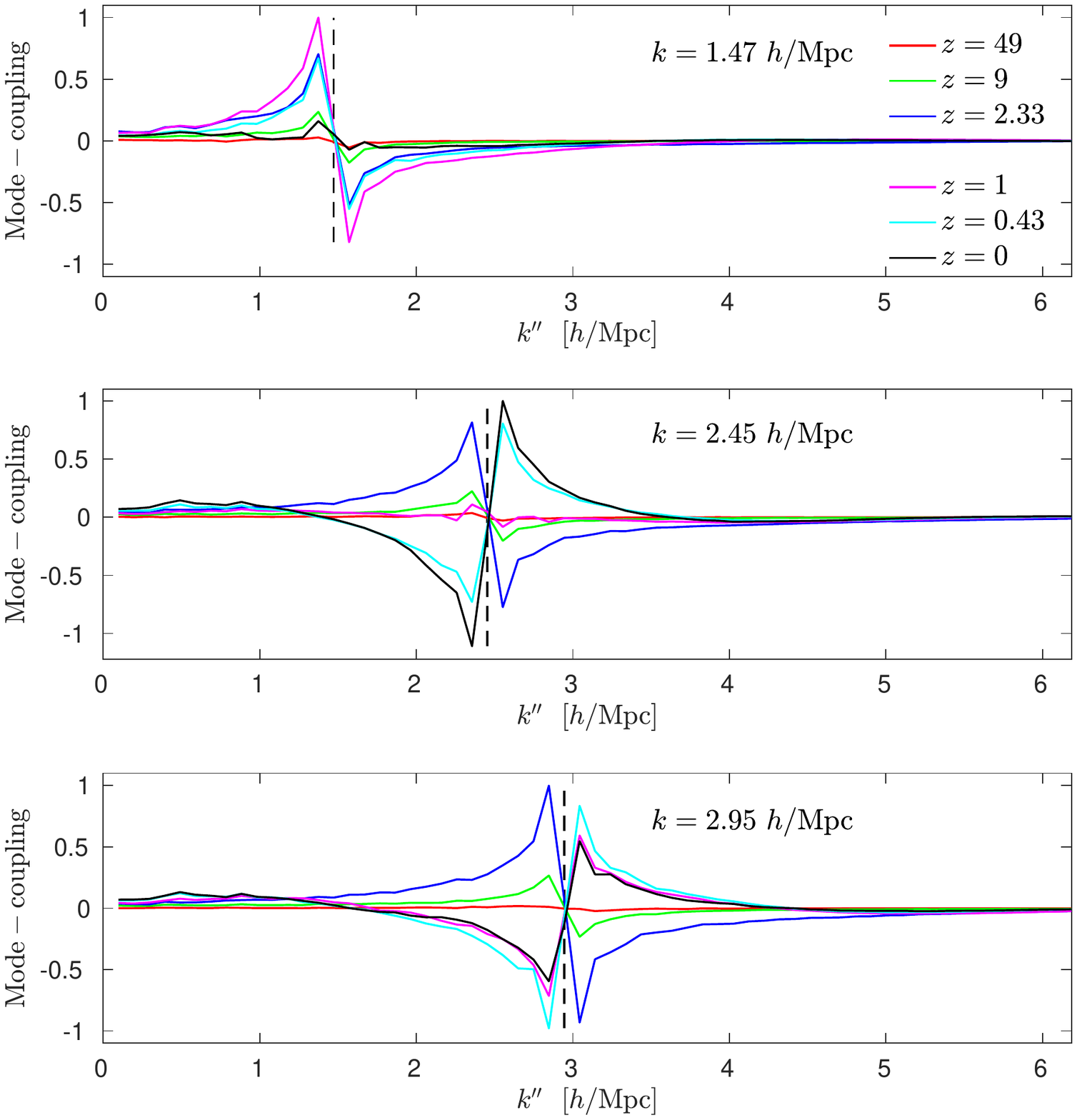}\end{center}  \vspace*{-3.3cm}
                              \caption{The figure shows the amount of mode coupling in the continuity equation, $i_{\theta\delta}^\theta(k'';k)$ between $\theta$ and $\delta$, as a function of $k''$ for specific values of $k$. The figure is based on simulation B, see Table~\ref{table:nbody_sims}.}
                              \label{fig:slices50}
\end{figure}

Figs.~\ref{fig:slices46} and \ref{fig:slices50} show the redshift evolution of $i_{\theta\delta}^\theta(k'';k)$ divided by the square of the linear growth factor.\footnote{In practice we approximate the square of the growth factor by $\sqrt{P_\delta(3k_f)P_\theta(3k_f)}$ for the continuity equation and $P_\theta(3k_f)$ for the Euler equation, where $k_f$ is the fundamental frequency of the simulation volume.}

Focusing on Fig.~\ref{fig:slices46}, it can be seen that there is a particular pattern that builds up, and that a given mode mostly couples to density perturbations which have roughly the same magnitude of the wavenumber. It can also clearly be seen that {\it statistically, power is constantly moved from larger to smaller scales}.

From Fig.~\ref{fig:slices46} it can be seen that the absolute amplitude of mode coupling, divided by the linear growth factor squared, grows as a function of redshift. For $k=0.74 h /{\rm Mpc}$ the mode coupling levels off to a constant shape and amplitude for $z \lesssim 1$. For this redshift range, the perturbations in this wavenumber can therefore be calculated with the linear theory growth rate multiplied with a factor that is time-$independent$.

\begin{figure}[h!]     \vspace*{-3.5cm}\begin{center}\hspace*{-0.5cm}\includegraphics[width=16cm]{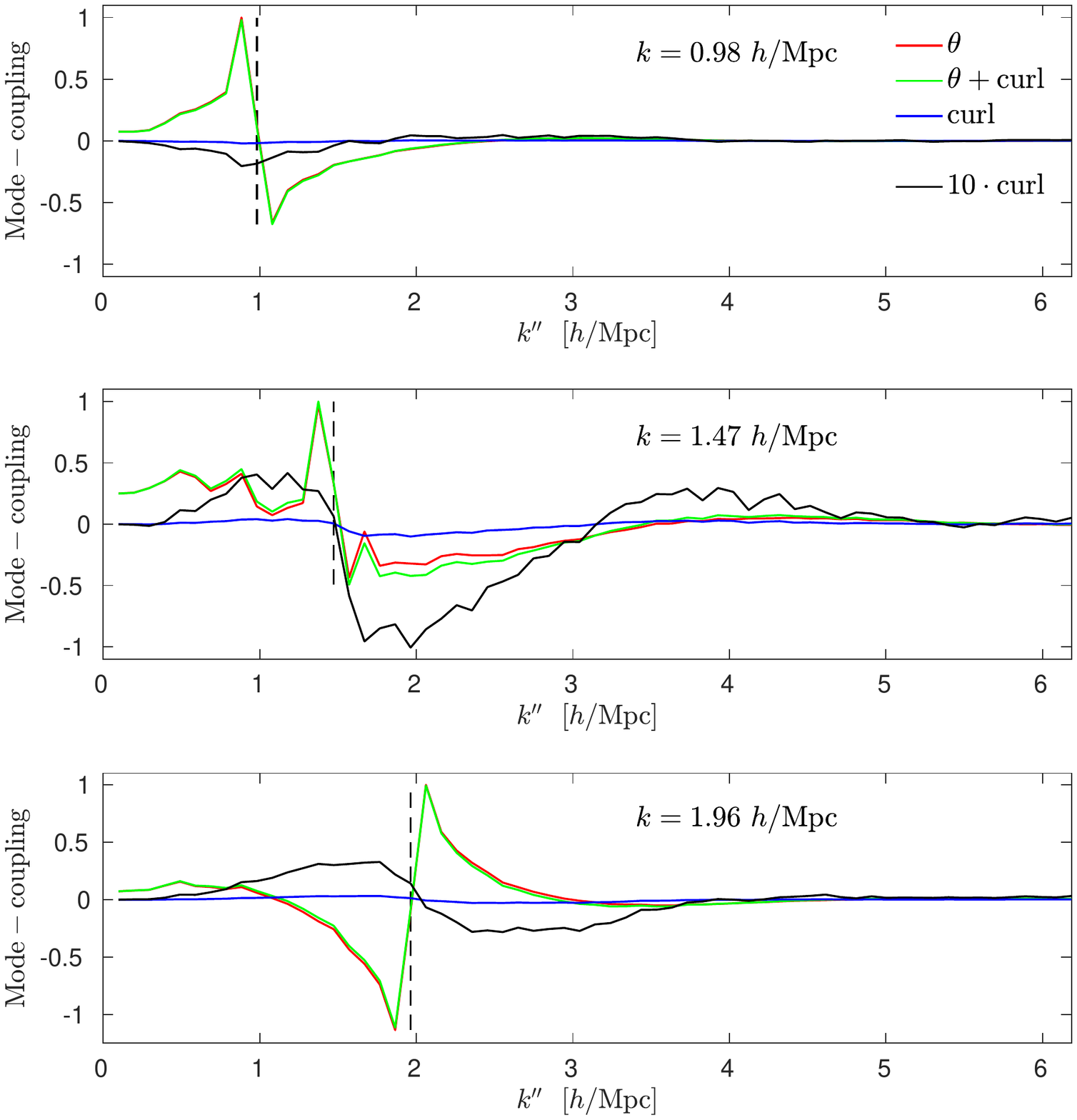}\end{center}  \vspace*{-3.3cm}
	         	     \caption{The figure shows the amount of mode coupling coming from the density-curl term, $i_{w\delta}^\theta(k'';k)$ (blue line), in the continuity equation at $z=0$ (data from simulation B). The red line shows $i_{\theta\delta}^\theta(k'';k)$ and the green line gives their sum.}
		             \label{fig:curl_z0}
\end{figure}

\begin{figure}[h!]     \vspace*{-3.5cm} \begin{center}\hspace*{-0.5cm}\includegraphics[width=17cm]{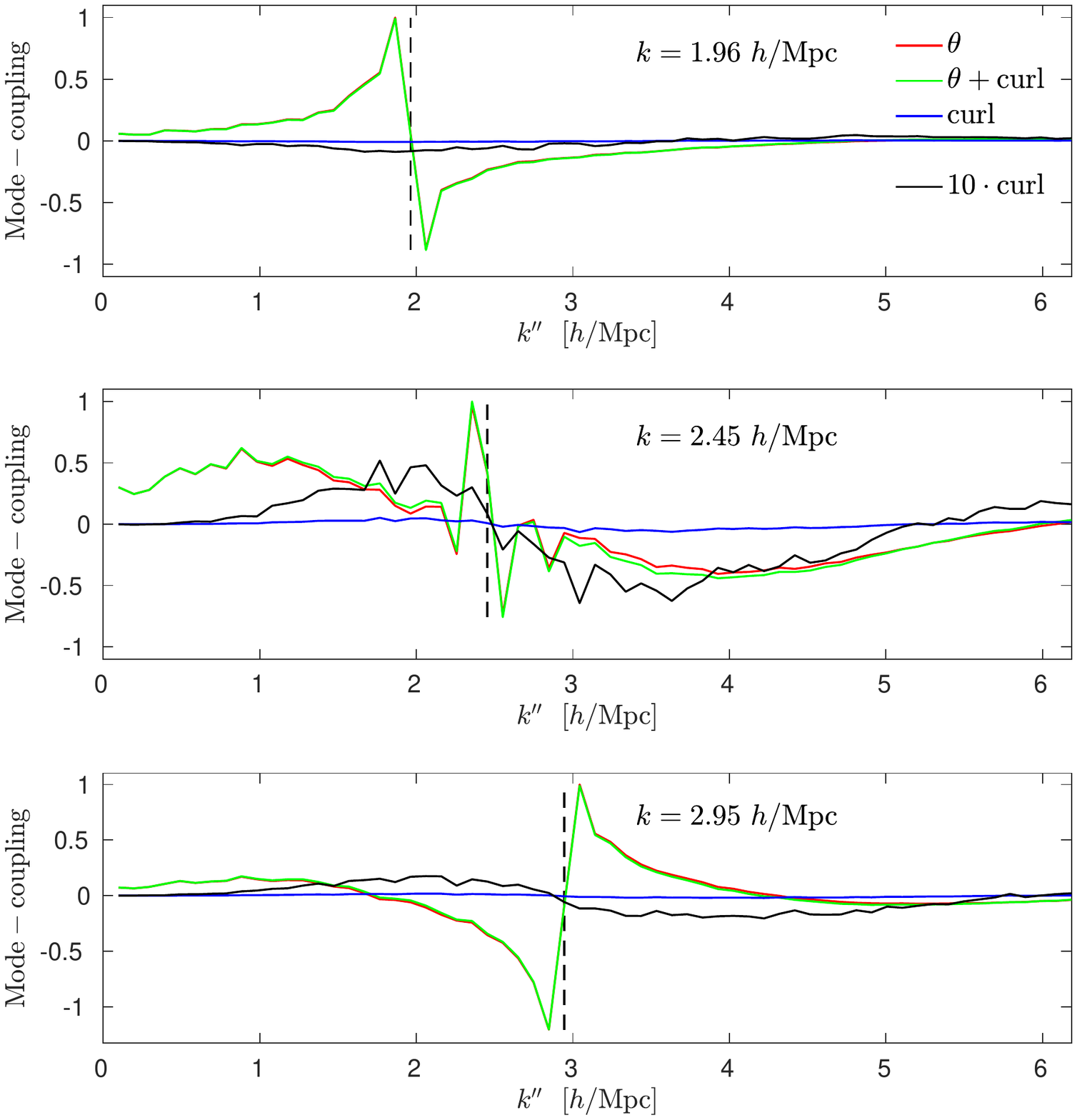}\end{center}  \vspace*{-3.3cm}
               	             \caption{Same as for Fig.~\ref{fig:curl_z0}, but at $z=1$.}			    
                             \label{fig:curl_z1}
\end{figure}

For $k > 1h/{\rm Mpc}$, see Fig.~\ref{fig:slices50}, the direction of power transfer becomes redshift dependent. For $k\sim 1.5h/{\rm Mpc}$, the amplitude begins to decrease for $z\lesssim 1$, while for $k$ larger than $\sim 2h/{\rm Mpc}$ the pattern has changed sign at low redshift. At $k\sim 3h/{\rm Mpc}$ the pattern first changes sign, and then decreases its amplitude once again. Notice that this sign change is local, in the sense that it does not happen for scales where $k''$ and $k$ are disparate.

The scale and redshift where this pattern shows up, matches the scale of virialisation. The change in sign can be explained by the fact that during infall smaller halos merge and transfer power from smaller to larger scales. Furthermore, fluctuations can get stretched by tidal forces as they fall into larger structures.

\begin{figure}[h!]
                            \vspace*{-7.2cm}\begin{center}\hspace*{0.5cm}\includegraphics[width=14.0cm]{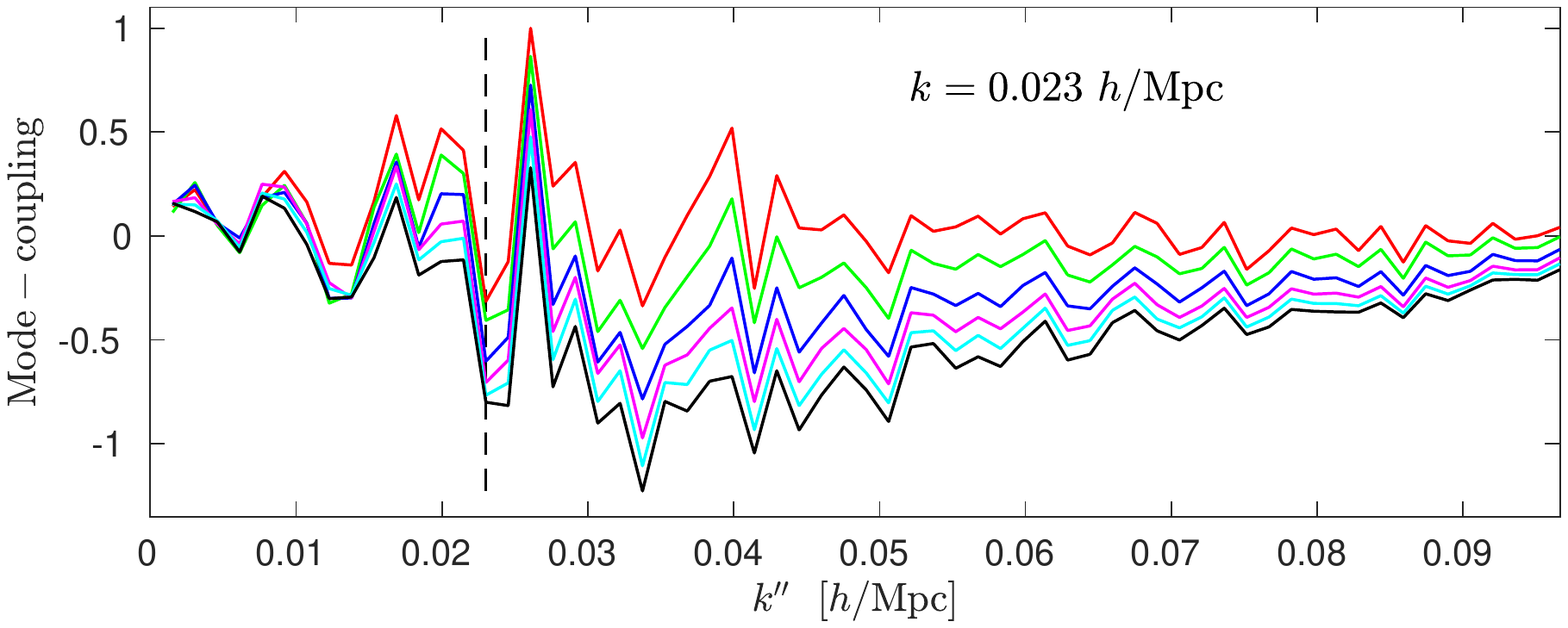}\end{center}\vspace*{-6.8cm}
                             \vspace*{-7.0cm}\begin{center}\hspace*{0.5cm}\includegraphics[width=14.0cm]{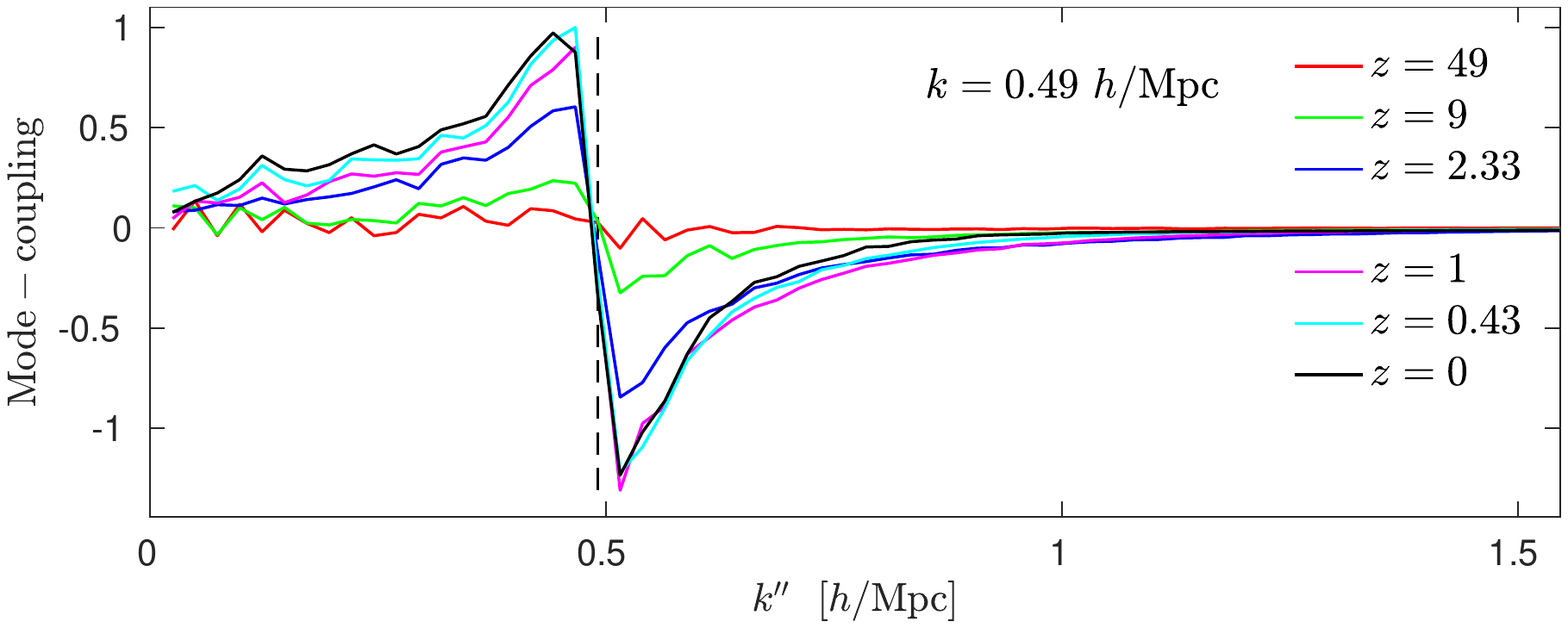}\end{center}\vspace*{-6.8cm}
                             \vspace*{-7.0cm}\begin{center}\hspace*{0.5cm}\includegraphics[width=14.0cm]{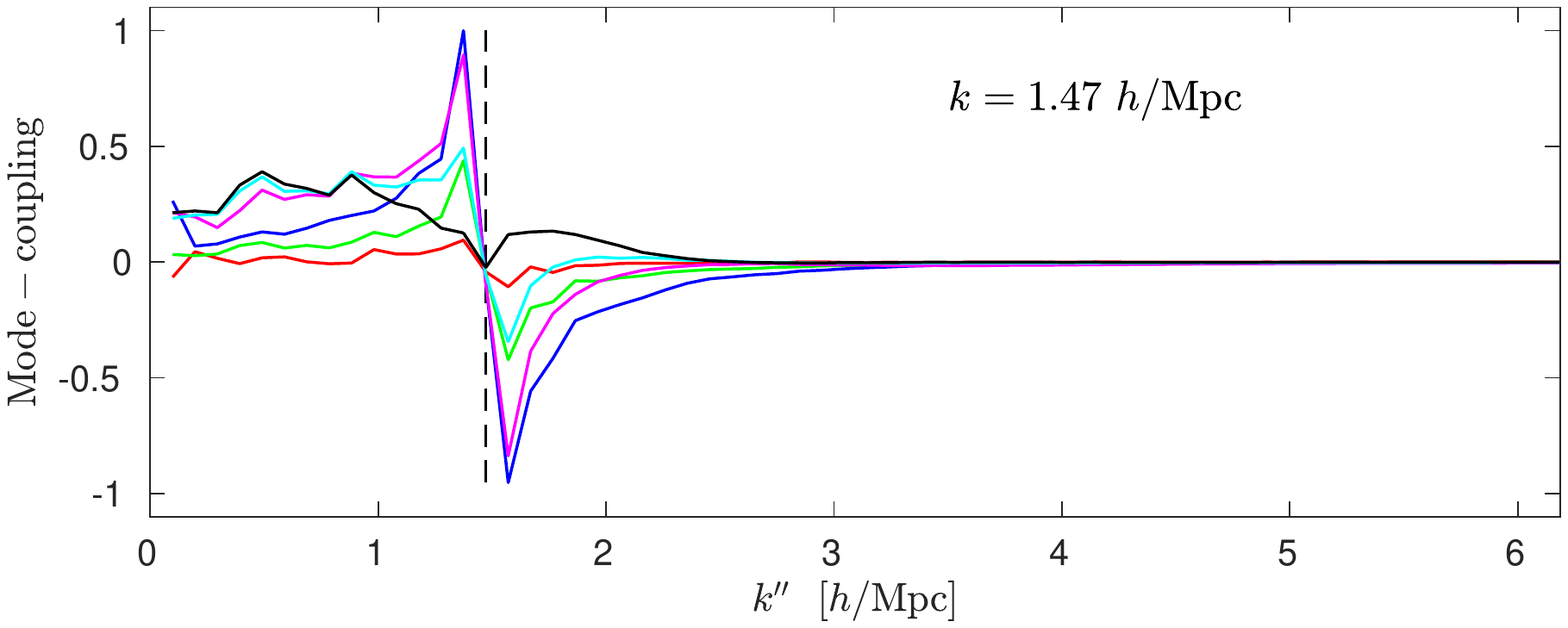}\end{center}\vspace*{-6.8cm}
                             \vspace*{-7.0cm}\begin{center}\hspace*{0.5cm}\includegraphics[width=14.0cm]{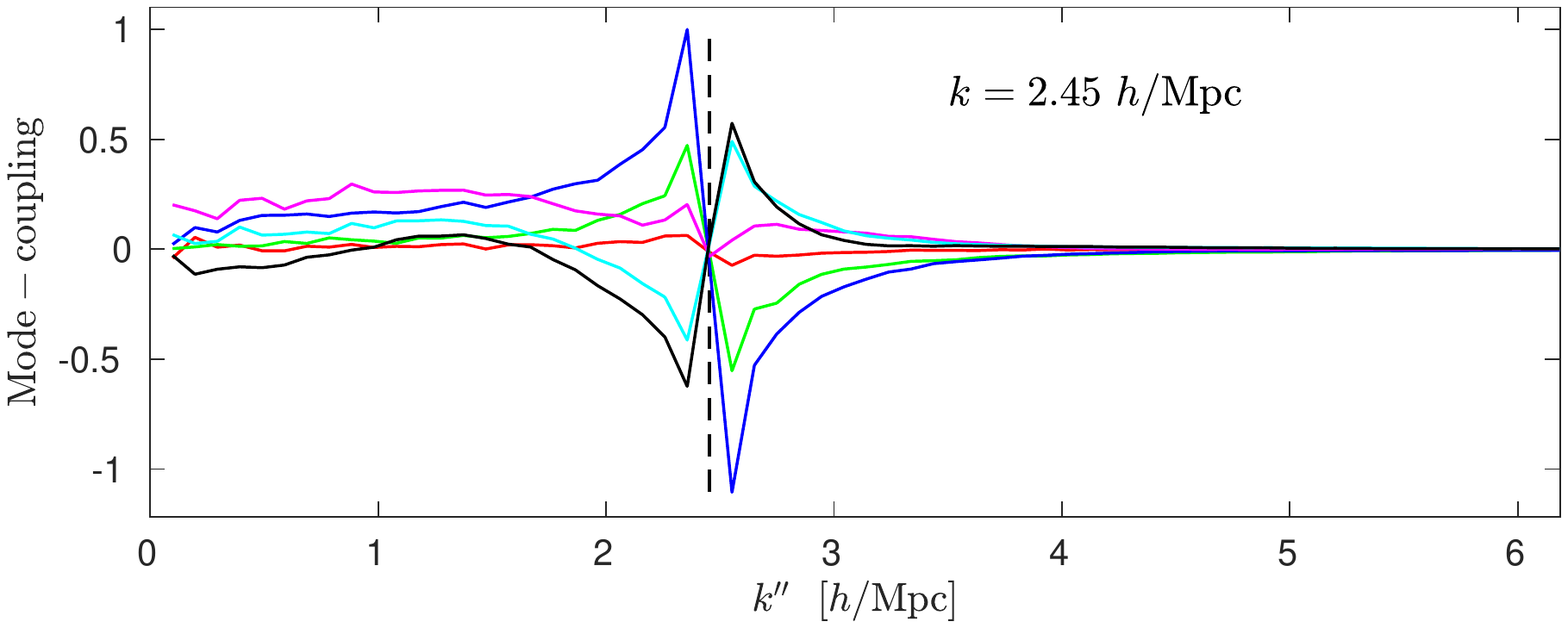}\end{center}\vspace*{-6.8cm}
                             \caption{The figure shows the amount of $\theta$-$\theta$ mode coupling in the Euler equation, as a function of $k''$ for specific values of $k$, i.e. $i_{\theta\theta}^\delta$. The upper panel is based on data from simulation F, panel number 2 from simulation D, and the lower two panels from simulation B.}
                             \label{fig:euler} 
\end{figure}

It is worth stressing that this pattern in Fourier space is transient and that the phenomena is followed by the stable clustering regime.

Instead of binning in $k''$ one could bin in $k'$. But since $k'$ is not the argument of the variable driving the physics in the continuity equation, namely $\delta$, this binning method does not give significant physical insight.

If the binning is instead performed in the angle between $\mathbf{k}$ and $\mathbf{k''}$, one finds that the largest contributions to the mode coupling integral come from small angles. From a variance point of view, this can easily be understood from the fact that, in this case, $k'$ is reduced, thereby increasing the expected absolute value of the terms $\theta(\mathbf{k'})$ and $\mathbf{k}\cdot \mathbf{k'}/k'^2$. 

Figs.~\ref{fig:curl_z0} and \ref{fig:curl_z1} show the effect of curl at $z=0$ and $z=1$, respectively, calculated from the simulation with a $64 {\rm Mpc} / h$ box. It can be seen that the effect of curl is limited, and matters the most for scales while they are in the process of virialisation.

From the figures it can be seen that before virialisation, the effect of curl is to reduce non-linear structure formation, since the effect of curl is negative in the upper panels at both redshifts. During virialisation, middle panels, the pattern is more complex. But after virialisation, lower panels, the effect of curl is opposite the effect of the divergence, i.e., the curl works in the opposite direction of the total flow of power.

The effect of curl appears to be small since we show $i_{w\delta}^\theta$, i.e.~the curl contribution to the mode coupling integral when correlating with the divergence. One should not expect curl to correlate significantly with the divergence, but to facilitate a comparison as a function of $k''$ we must display $i_{\theta\delta}^\theta$ and $i_{w\delta}^\theta$, i.e.~both quantities must be correlated with the same fluid variable.

Fig.~\ref{fig:euler} shows the evolution of the dominant mode coupling term in the Euler equation, namely $i_{\theta\theta}^\delta(k'';k)$. The top panel shows the time-evolution for $k=0.023 h /{\rm Mpc}$, which is around the scale where the linear and non-linear divergence power spectra begin to diverge today. Despite the presence of noise, it can clearly be seen that the mode coupling leads to divergence being 'moved' towards smaller scales, i.e.~with the interpretation that the divergence decreases.

The panel with $k=0.49 h/{\rm Mpc}$ shows a pattern that is reminiscent of the one found from the $i_{\theta\delta}^\theta$ integrand in the continuity equation, except for the fact that the negative amplitude is larger in absolute terms than the positive amplitude. Again, this accounts for the fact that the non-linear divergence is smaller than the linear one. The difference between the mode coupling integrals in the continuity and Euler equations at semi-linear scales is the presence of the extra factor of $-1$ in the kernel of $i_{\theta\theta}^\delta$ as compared to $i_{\theta\delta}^\theta$ (the $k''^{-2}$ factor only leads to an even larger difference, compare Eqs.~(\ref{eq:I_td}) and (\ref{eq:I_tt}) with the substitution $\mathbf{k''=k-k'}$). It is this extra factor that accounts for the suppression of the velocity divergence in non-linear theory as compared to linear theory. 

The two lower panels in Fig.~\ref{fig:euler} show the same qualitative behaviour as was seen from the continuity equation. Interestingly, as virialisation occurs, the sum over $i_{\theta\theta}^\delta$ tends towards positive values,  so that the non-linear divergence power spectrum grows faster than its linear counterpart. The physical driver is the $k^2 \psi(\mathbf{k})$ term at very non-linear scales, and the effect can be seen as a change in slope for $P_\theta$ at low redshift and $k\gtrsim 1$ (see Fig~\ref{fig:power}).

We have not added the $\theta$-curl and curl-curl terms to the $\theta$-$\theta$ term, since the latter term is the dominant term, except on very small scales, $k\gtrsim 2-5 h/{\rm Mpc}$. We have likewise not added the velocity dispersion term, $q_\theta$, but as indicated in \cite{Pueblas:2008uv}, $q_\theta$ might become relevant for $k \gtrsim 1 h/{\rm Mpc}$ at $z=0$. But we stress that adding these sub-dominant terms, do not affect the qualitative conclusions presented in this paper. 

To sum up, at $z=0$ the scale $k\sim 1.5h/{\rm Mpc}$ (length scale $\sim 6~ {\rm Mpc}$) stands out as the scale where the $\delta$ and $\theta$ power spectra change their slope and where the vorticity power spectrum attains its maximum. It is also the scale where the mode coupling pattern in Fig.~\ref{fig:slices50}, upper panel, and Fig.~\ref{fig:euler}, third panel, becomes roughly featureless. It is likewise the scale where the effect of vorticity is largest, see Fig.~\ref{fig:curl_z0}, middle panel. The above considerations are also valid at $z=1$, if one translates to a length scale of $\sim 4~{\rm Mpc}$ corresponding to $k\sim 2.5 h/{\rm Mpc}$.

%%%%%%%%%%%%%%%%%%%%%%%%%%%%%%%%%%%%%%%%%%%%%%%%%%%%%%%%%%%%%%%%%%%%%%%%%%%%%%%%%%%%%%%%%%%%%%% 
\section{Conclusions}\label{sec:conclusions}

We have presented a novel way of studying non-linear structure formation, namely by imaging the fully non-linear mode coupling integrals appearing in the Fourier space continuity and Euler equations using $N$-body simulations.
This method allows for a direct reconstruction of the coupling between any two wavenumbers and can be easily visualised.

In accordance with expectations we find that in the linear regime there is no mode coupling. In the mildly non-linear regime the mode coupling integrals are dominated by transfer of power from large to small scales, primarily coupling modes which are close in $k$-space. This behaviour corresponds to the infall regime where structure formation is dominated by the initial collapse of structures, prior to the onset of virialisation.

On smaller scales (corresponding to $k \sim 2-3 \, h$/Mpc at $z=0$) where virialisation sets in and the stable clustering regime is entered, this simple behaviour becomes much more complex. Around this scale vorticity starts to build up and structure formation slows down. In the mode coupling integrals this can be seen as a temporary reversal of the power transfer (lack of small scale resolution in our simulations prevents us from studying this effect in the extremely small scale limit). This change in behaviour sets in exactly where the curl contribution to the velocity power spectrum becomes comparable to the divergence part, and qualitatively the effect can be thought of as the build-up of pressure once virialisation sets in and the local velocity tensor is isotropised.

The results presented here are a step towards understanding the behaviour of the mode coupling integrals in the fully non-linear regime and our results can potentially be used to calibrate and improve semi-analytic models of structure formation, either by having these models calculate the statistics presented in this paper, or by extending the tools presented in this paper to calculate quantities which naturally arise in various perturbation theory models.
In this work we have not investigated how the mode coupling function shown in Figs.~\ref{fig:slices46} and \ref{fig:slices50} depends on the cosmological model used, but this will be the focus of a future study. If the functional form turns out to be close to universal it will be extremely useful for devising new, semi-analytic models of non-linear structure formation based on the Fourier space fluid equations. Since we have not included anisotropic stress in our calculations, calibration of semi-analytic models to our results cannot be performed reliably below the virialization scale, $k\gtrsim 1 h/ {\rm Mpc}$ at $z\sim 0$.

\section*{Acknowledgements}
We acknowledge computing resources from the Danish Center for Scientific Computing (DCSC), and thank Y.~Y.~Y.~Wong and T.~Tram for comments on the manuscript. This work was supported by the Villum Foundation.

%%%%%%%%%%%%%%%%%%%%%%%%%%%%%%%%%%%%%%%%%%%%%%%%%%%%%%%%%%%%%%%%%%%%%%%%%%%%%%%%%%%%%%%%%%%%%%%
%%%%%%%%%%%%%%%%%%%%%%%%%%%%%%%

\bibliographystyle{utcaps}

\begin{thebibliography}{10}

\bibitem{Peloso:2016qdr}
  M.~Peloso and M.~Pietroni,
  ``Galilean invariant resummation schemes of cosmological perturbations,''
  [arXiv:1609.06624 [astro-ph.CO]].
  %%CITATION = ARXIV:1609.06624;%%

%\cite{Floerchinger:2016hja}
\bibitem{Floerchinger:2016hja}
  S.~Floerchinger, M.~Garny, N.~Tetradis and U.~A.~Wiedemann,
  ``Renormalization-group flow of the effective action of cosmological large-scale structures,''
  [arXiv:1607.03453 [astro-ph.CO]].
  %%CITATION = ARXIV:1607.03453;%%
  %3 citations counted in INSPIRE as of 20 Jan 2017

%\cite{Senatore:2014via}
\bibitem{Senatore:2014via}
  L.~Senatore and M.~Zaldarriaga,
  ``The IR-resummed Effective Field Theory of Large Scale Structures,''
  JCAP {\bf 1502} (2015) no.02,  013
%  doi:10.1088/1475-7516/2015/02/013
  [arXiv:1404.5954 [astro-ph.CO]].
  %%CITATION = doi:10.1088/1475-7516/2015/02/013;%%
  %51 citations counted in INSPIRE as of 20 Jan 2017

%\cite{Carroll:2013oxa}
\bibitem{Carroll:2013oxa}
  S.~M.~Carroll, S.~Leichenauer and J.~Pollack,
  ``Consistent effective theory of long-wavelength cosmological perturbations,''
  Phys.\ Rev.\ D {\bf 90} (2014) no.2,  023518
%  doi:10.1103/PhysRevD.90.023518
  [arXiv:1310.2920 [hep-th]].
  %%CITATION = doi:10.1103/PhysRevD.90.023518;%%
  %49 citations counted in INSPIRE as of 20 Jan 2017

%\cite{Carrasco:2012cv}
\bibitem{Carrasco:2012cv}
  J.~J.~M.~Carrasco, M.~P.~Hertzberg and L.~Senatore,
  ``The Effective Field Theory of Cosmological Large Scale Structures,''
  JHEP {\bf 1209} (2012) 082
%  doi:10.1007/JHEP09(2012)082
  [arXiv:1206.2926 [astro-ph.CO]].
  %%CITATION = doi:10.1007/JHEP09(2012)082;%%
  %132 citations counted in INSPIRE as of 20 Jan 2017

%\cite{Bernardeau:2008fa}
\bibitem{Bernardeau:2008fa}
  F.~Bernardeau, M.~Crocce and R.~Scoccimarro,
  ``Multi-Point Propagators in Cosmological Gravitational Instability,''
  Phys.\ Rev.\ D {\bf 78} (2008) 103521
%  doi:10.1103/PhysRevD.78.103521
  [arXiv:0806.2334 [astro-ph]].
  %%CITATION = doi:10.1103/PhysRevD.78.103521;%%
  %101 citations counted in INSPIRE as of 20 Jan 2017

%\cite{Matsubara:2007wj}
\bibitem{Matsubara:2007wj}
  T.~Matsubara,
  ``Resumming Cosmological Perturbations via the Lagrangian Picture: One-loop Results in Real Space and in Redshift Space,''
  Phys.\ Rev.\ D {\bf 77} (2008) 063530
%  doi:10.1103/PhysRevD.77.063530
  [arXiv:0711.2521 [astro-ph]].
  %%CITATION = doi:10.1103/PhysRevD.77.063530;%%
  %236 citations counted in INSPIRE as of 20 Jan 2017

%\cite{Pietroni:2008jx}
\bibitem{Pietroni:2008jx}
  M.~Pietroni,
  ``Flowing with Time: a New Approach to Nonlinear Cosmological Perturbations,''
  JCAP {\bf 0810} (2008) 036
%  doi:10.1088/1475-7516/2008/10/036
  [arXiv:0806.0971 [astro-ph]].
  %%CITATION = doi:10.1088/1475-7516/2008/10/036;%%
  %136 citations counted in INSPIRE as of 20 Jan 2017

%\cite{Lesgourgues:2009am}
\bibitem{Lesgourgues:2009am}
  J.~Lesgourgues, S.~Matarrese, M.~Pietroni and A.~Riotto,
  ``Non-linear Power Spectrum including Massive Neutrinos: the Time-RG Flow Approach,''
  JCAP {\bf 0906} (2009) 017
%  doi:10.1088/1475-7516/2009/06/017
  [arXiv:0901.4550 [astro-ph.CO]].
  %%CITATION = doi:10.1088/1475-7516/2009/06/017;%%
  %64 citations counted in INSPIRE as of 20 Jan 2017

%\cite{Matarrese:2007wc}
\bibitem{Matarrese:2007wc}
  S.~Matarrese and M.~Pietroni,
  ``Resumming Cosmic Perturbations,''
  JCAP {\bf 0706} (2007) 026
%  doi:10.1088/1475-7516/2007/06/026
  [astro-ph/0703563].
  %%CITATION = doi:10.1088/1475-7516/2007/06/026;%%
  %117 citations counted in INSPIRE as of 20 Jan 2017

%\cite{Crocce:2005xy}
\bibitem{Crocce:2005xy}
  M.~Crocce and R.~Scoccimarro,
  ``Renormalized cosmological perturbation theory,''
  Phys.\ Rev.\ D {\bf 73} (2006) 063519
%  doi:10.1103/PhysRevD.73.063519
  [astro-ph/0509418].
  %%CITATION = doi:10.1103/PhysRevD.73.063519;%%
  %301 citations counted in INSPIRE as of 20 Jan 2017

\bibitem{Blas:2013aba}
  D.~Blas, M.~Garny and T.~Konstandin,
  ``Cosmological perturbation theory at three-loop order,''
  JCAP {\bf 1401} (2014) no.01,  010
%  doi:10.1088/1475-7516/2014/01/010
  [arXiv:1309.3308 [astro-ph.CO]].
  %%CITATION = doi:10.1088/1475-7516/2014/01/010;%%
  %29 citations counted in INSPIRE as of 20 Jan 2017

%\cite{Carlson:2009it}
\bibitem{Carlson:2009it}
  J.~Carlson, M.~White and N.~Padmanabhan,
  ``A critical look at cosmological perturbation theory techniques,''
  Phys.\ Rev.\ D {\bf 80} (2009) 043531
%  doi:10.1103/PhysRevD.80.043531
  [arXiv:0905.0479 [astro-ph.CO]].
  %%CITATION = doi:10.1103/PhysRevD.80.043531;%%
  %125 citations counted in INSPIRE as of 20 Jan 2017

%\cite{Carrasco:2013sva}
\bibitem{Carrasco:2013sva}
  J.~J.~M.~Carrasco, S.~Foreman, D.~Green and L.~Senatore,
  ``The 2-loop matter power spectrum and the IR-safe integrand,''
  JCAP {\bf 1407} (2014) 056
%  doi:10.1088/1475-7516/2014/07/056
  [arXiv:1304.4946 [astro-ph.CO]].
  %%CITATION = doi:10.1088/1475-7516/2014/07/056;%%
  %47 citations counted in INSPIRE as of 20 Jan 2017

%\cite{Rampf:2016wom}
\bibitem{Rampf:2016wom}
  C.~Rampf, E.~Villa, D.~Bertacca and M.~Bruni,
  ``Lagrangian theory for cosmic structure formation with vorticity: Newtonian and post-Friedmann approximations,''
  Phys.\ Rev.\ D {\bf 94} (2016) no.8,  083515
%  doi:10.1103/PhysRevD.94.083515
  [arXiv:1607.05226 [gr-qc]].
  %%CITATION = doi:10.1103/PhysRevD.94.083515;%%
  %1 citations counted in INSPIRE as of 20 Jan 2017

%\cite{Villa:2015ppa}
\bibitem{Villa:2015ppa}
  E.~Villa and C.~Rampf,
  ``Relativistic perturbations in $\Lambda$CDM: Eulerian \& Lagrangian approaches,''
  JCAP {\bf 1601} (2016) no.01,  030
%  doi:10.1088/1475-7516/2016/01/030
  [arXiv:1505.04782 [gr-qc]].
  %%CITATION = doi:10.1088/1475-7516/2016/01/030;%%
  %8 citations counted in INSPIRE as of 20 Jan 2017

%\cite{Rampf:2012xb}
\bibitem{Rampf:2012xb}
  C.~Rampf and Y.~Y.~Y.~Wong,
  ``Lagrangian perturbations and the matter bispectrum II: the resummed one-loop correction to the matter bispectrum,''
  JCAP {\bf 1206} (2012) 018
%  doi:10.1088/1475-7516/2012/06/018
  [arXiv:1203.4261 [astro-ph.CO]].
  %%CITATION = doi:10.1088/1475-7516/2012/06/018;%%
  %17 citations counted in INSPIRE as of 20 Jan 2017

%\cite{Scoccimarro:1996se}
\bibitem{Scoccimarro:1996se}
  R.~Scoccimarro and J.~Frieman,
  ``Loop corrections in nonlinear cosmological perturbation theory 2. Two point statistics and selfsimilarity,''
  Astrophys.\ J.\  {\bf 473} (1996) 620
%  doi:10.1086/178177
  [astro-ph/9602070].
  %%CITATION = doi:10.1086/178177;%%
  %82 citations counted in INSPIRE as of 20 Jan 2017

%\cite{Smith:2002dz}
\bibitem{Smith:2002dz}
  R.~E.~Smith {\it et al.} [VIRGO Consortium Collaboration],
  ``Stable clustering, the halo model and nonlinear cosmological power spectra,''
  Mon.\ Not.\ Roy.\ Astron.\ Soc.\  {\bf 341} (2003) 1311
%  doi:10.1046/j.1365-8711.2003.06503.x
  [astro-ph/0207664].
  %%CITATION = doi:10.1046/j.1365-8711.2003.06503.x;%%
  %1123 citations counted in INSPIRE as of 20 Dec 2016

%\cite{Takahashi:2012em}
\bibitem{Takahashi:2012em}
  R.~Takahashi, M.~Sato, T.~Nishimichi, A.~Taruya and M.~Oguri,
  ``Revising the Halofit Model for the Nonlinear Matter Power Spectrum,''
  Astrophys.\ J.\  {\bf 761} (2012) 152
  %doi:10.1088/0004-637X/761/2/152
  [arXiv:1208.2701 [astro-ph.CO]].
  %%CITATION = doi:10.1088/0004-637X/761/2/152;%%
  %211 citations counted in INSPIRE as of 23 Jan 2017
	
%\cite{Schneider:2015yka}
\bibitem{Schneider:2015yka}
  A.~Schneider {\it et al.},
  ``Matter power spectrum and the challenge of percent accuracy,''
  JCAP {\bf 1604} (2016) no.04,  047
%  doi:10.1088/1475-7516/2016/04/047
  [arXiv:1503.05920 [astro-ph.CO]].
  %%CITATION = doi:10.1088/1475-7516/2016/04/047;%%
  %27 citations counted in INSPIRE as of 25 Oct 2016  

  %\cite{Pueblas:2008uv}
\bibitem{Pueblas:2008uv}
  S.~Pueblas and R.~Scoccimarro,
  ``Generation of Vorticity and Velocity Dispersion by Orbit Crossing,''
  Phys.\ Rev.\ D {\bf 80} (2009) 043504
%  doi:10.1103/PhysRevD.80.043504
  [arXiv:0809.4606 [astro-ph]].
  %%CITATION = doi:10.1103/PhysRevD.80.043504;%%
  %80 citations counted in INSPIRE as of 20 Oct 2016

%\cite{Nishimichi:2014rra}
\bibitem{Nishimichi:2014rra}
  T.~Nishimichi, F.~Bernardeau and A.~Taruya,
  ``Response function of the large-scale structure of the universe to the small scale inhomogeneities,''
  Phys.\ Lett.\ B {\bf 762} (2016) 247
%  doi:10.1016/j.physletb.2016.09.035
  [arXiv:1411.2970 [astro-ph.CO]].
  %%CITATION = doi:10.1016/j.physletb.2016.09.035;%%
  %9 citations counted in INSPIRE as of 21 Oct 2016
  
  %\cite{Ma:1995ey}
\bibitem{Ma:1995ey}
  C.~P.~Ma and E.~Bertschinger,
  ``Cosmological perturbation theory in the synchronous and conformal Newtonian gauges,''
  Astrophys.\ J.\  {\bf 455} (1995) 7
%  doi:10.1086/176550
  [astro-ph/9506072].
  %%CITATION = doi:10.1086/176550;%%
  %1047 citations counted in INSPIRE as of 20 Oct 2016
  
  %\cite{Jain:1993yk}
\bibitem{Jain:1993yk}
  B.~Jain and E.~Bertschinger,
  ``Nonlinear effects due to the coupling of long wave modes,''
  [astro-ph/9309040].
  %%CITATION = ASTRO-PH/9309040;%%
  
    %\cite{Jain:1993jh}
\bibitem{Jain:1993jh}
  B.~Jain and E.~Bertschinger,
 ``Second order power spectrum and nonlinear evolution at high redshift,''
  Astrophys.\ J.\  {\bf 431} (1994) 495
%  doi:10.1086/174502
  [astro-ph/9311070].
  %%CITATION = doi:10.1086/174502;%%
  %186 citations counted in INSPIRE as of 20 Oct 2016
  
  %\cite{Audren:2011ne}
\bibitem{Audren:2011ne}
  B.~Audren and J.~Lesgourgues,
  ``Non-linear matter power spectrum from Time Renormalisation Group: efficient computation and comparison with one-loop,''
  JCAP {\bf 1110} (2011) 037
%  doi:10.1088/1475-7516/2011/10/037
  [arXiv:1106.2607 [astro-ph.CO]].
  %%CITATION = doi:10.1088/1475-7516/2011/10/037;%%
  %13 citations counted in INSPIRE as of 20 Oct 2016
  
%\cite{Bernardeau:2001qr}
\bibitem{Bernardeau:2001qr}
  F.~Bernardeau, S.~Colombi, E.~Gaztanaga and R.~Scoccimarro,
  ``Large scale structure of the universe and cosmological perturbation theory,''
  Phys.\ Rept.\  {\bf 367} (2002) 1
%  doi:10.1016/S0370-1573(02)00135-7
  [astro-ph/0112551].
  %%CITATION = doi:10.1016/S0370-1573(02)00135-7;%%
  %713 citations counted in INSPIRE as of 30 Oct 2016
  
    %\cite{Ade:2015xua}
\bibitem{Ade:2015xua}
  P.~A.~R.~Ade {\it et al.} [Planck Collaboration],
  ``Planck 2015 results. XIII. Cosmological parameters,''
  Astron.\ Astrophys.\  {\bf 594} (2016) A13
  %doi:10.1051/0004-6361/201525830
  [arXiv:1502.01589 [astro-ph.CO]].
  %%CITATION = doi:10.1051/0004-6361/201525830;%%
  %2617 citations counted in INSPIRE as of 20 Dec 2016

  %\cite{Lewis:2002ah}
\bibitem{Lewis:2002ah}
  A.~Lewis and S.~Bridle,
  ``Cosmological parameters from CMB and other data: A Monte Carlo approach,''
  Phys.\ Rev.\ D {\bf 66} (2002) 103511
%  doi:10.1103/PhysRevD.66.103511
  [astro-ph/0205436].
  %%CITATION = doi:10.1103/PhysRevD.66.103511;%%
  %1883 citations counted in INSPIRE as of 11 Dec 2016
  
%\cite{Brandbyge:2008js}
\bibitem{Brandbyge:2008js}
  J.~Brandbyge and S.~Hannestad,
  ``Grid Based Linear Neutrino Perturbations in Cosmological N-body Simulations,''
  JCAP {\bf 0905} (2009) 002
%  doi:10.1088/1475-7516/2009/05/002
  [arXiv:0812.3149 [astro-ph]].
  %%CITATION = doi:10.1088/1475-7516/2009/05/002;%%
  
  %\cite{Springel:2005mi}
\bibitem{Springel:2005mi}
  V.~Springel,
  ``The Cosmological simulation code GADGET-2,''
  Mon.\ Not.\ Roy.\ Astron.\ Soc.\  {\bf 364} (2005) 1105
%  doi:10.1111/j.1365-2966.2005.09655.x
  [astro-ph/0505010].
  %%CITATION = doi:10.1111/j.1365-2966.2005.09655.x;%%
  %2398 citations counted in INSPIRE as of 11 Dec 2016

%\cite{Zeldovich:1969sb}
\bibitem{Zeldovich:1969sb}
  Y.~B.~Zeldovich,
  ``Gravitational instability: An Approximate theory for large density perturbations,''
  Astron.\ Astrophys.\  {\bf 5} (1970) 84.
  %%CITATION = AAEJA,5,84;%%
  %1134 citations counted in INSPIRE as of 11 Dec 2016
  
  %\cite{Crocce:2006ve}
\bibitem{Crocce:2006ve}
  M.~Crocce, S.~Pueblas and R.~Scoccimarro,
  ``Transients from Initial Conditions in Cosmological Simulations,''
  Mon.\ Not.\ Roy.\ Astron.\ Soc.\  {\bf 373} (2006) 369
%  doi:10.1111/j.1365-2966.2006.11040.x
  [astro-ph/0606505].
  %%CITATION = doi:10.1111/j.1365-2966.2006.11040.x;%%
  %272 citations counted in INSPIRE as of 11 Dec 2016
  
  %\cite{Bouchet:1994xp}
\bibitem{Bouchet:1994xp}
  F.~R.~Bouchet, S.~Colombi, E.~Hivon and R.~Juszkiewicz,
  ``Perturbative Lagrangian approach to gravitational instability,''
  Astron.\ Astrophys.\  {\bf 296} (1995) 575
  [astro-ph/9406013].
  %%CITATION = ASTRO-PH/9406013;%%
  %172 citations counted in INSPIRE as of 11 Dec 2016
  
  %\cite{Scoccimarro:1997gr}
\bibitem{Scoccimarro:1997gr}
  R.~Scoccimarro,
  ``Transients from initial conditions: a perturbative analysis,''
  Mon.\ Not.\ Roy.\ Astron.\ Soc.\  {\bf 299} (1998) 1097
%  doi:10.1046/j.1365-8711.1998.01845.x
  [astro-ph/9711187].
  %%CITATION = doi:10.1046/j.1365-8711.1998.01845.x;%%
  %163 citations counted in INSPIRE as of 11 Dec 2016

%\cite{Jennings:2012ej}
\bibitem{Jennings:2012ej}
  E.~Jennings,
  ``An improved model for the nonlinear velocity power spectrum,''
  Mon.\ Not.\ Roy.\ Astron.\ Soc.\  {\bf 427} (2012) L25
%  doi:10.1111/j.1745-3933.2012.01338.x
  [arXiv:1207.1439 [astro-ph.CO]].
  %%CITATION = doi:10.1111/j.1745-3933.2012.01338.x;%%
  %18 citations counted in INSPIRE as of 20 Oct 2016  

\bibitem{Monaghan:1985}
    J.~J.~Monaghan, J.~C.~Lattanzio, 
    ``A refined particle method for astrophysical problems,"
    Astron.\ Astrophys.\ 149, 135 (1985).  

%\cite{Cusin:2016zvu}
\bibitem{Cusin:2016zvu}
  G.~Cusin, V.~Tansella and R.~Durrer,
  ``Vorticity generation in the Universe: A perturbative approach,''
  arXiv:1612.00783 [astro-ph.CO].
  %%CITATION = ARXIV:1612.00783;%%  


\end{thebibliography}

\providecommand{\href}[2]{#2}\begingroup\raggedright\endgroup

\end{document}